\documentclass
[aps,prb,twocolumn,floatfix,english,showpacs,10pt]{revtex4-2}%
\usepackage{graphicx}
\usepackage{booktabs}
\usepackage{amsmath}
\usepackage{physics}
\usepackage{amssymb}
\usepackage{colordvi}
\usepackage{verbatim}
\usepackage{xcolor}
\usepackage{mathrsfs}
\usepackage{epsfig}
\usepackage{lipsum}
\usepackage{amsfonts}
\usepackage{makecell}
\usepackage{esint}
\usepackage{bm}
\usepackage[most]{tcolorbox}

\usepackage[unicode=true, breaklinks=false, pdfborder={0 0 1}, backref=false,
colorlinks=true, linkcolor=blue, urlcolor=blue, citecolor=blue]{hyperref}%
\providecommand{\U}[1]{\protect\rule{.1in}{.1in}}
\providecommand{\U}[1]{\protect\rule{.1in}{.1in}}
\setcitestyle{numbers,square}

\begin{document}

\title{Transverse and unidirectional spin pumping}

\author{Ping Li}
\thanks{These authors contributed equally to this work.}
\affiliation{School of Physics, Huazhong University of Science and Technology, Wuhan 430074, China}

\author{Chengyuan Cai}
\thanks{These authors contributed equally to this work.}
\affiliation{School of Physics, Huazhong University of Science and Technology, Wuhan 430074, China}
 
\author{Tao Yu}
\email{taoyuphy@hust.edu.cn}
\affiliation{School of Physics, Huazhong University of Science and Technology, Wuhan 430074, China}

\date{\today }

\begin{abstract}

Conventional spin pumping, driven by magnetization dynamics, is longitudinal since the pumped spin current flows normal to the interface between the ferromagnet and the conductor. We predict \textit{Hall-type/transverse} and \textit{unidirectional} spin pumping into conductors by near-field electromagnetic radiation emitted by, \textit{e.g.}, magnetization dynamics. The joint effect of the electric and magnetic fields results in a pure spin current flowing parallel to the interface, i.e., a Hall-type spin pumping, which is highly efficient due to the strong coupling to the electric field. Such a transverse spin current is unidirectional, with the spatial distribution controlled by the magnetization direction. Our finding reveals a robust approach for generating and manipulating spin currents in future low-dimensional spintronic and orbitronic devices.

\end{abstract}

\maketitle

\section{Introduction}

\begin{figure}[htp!]
    \centering
     \includegraphics[width=0.96\linewidth, clip, trim=2cm 4cm 1cm 4cm]{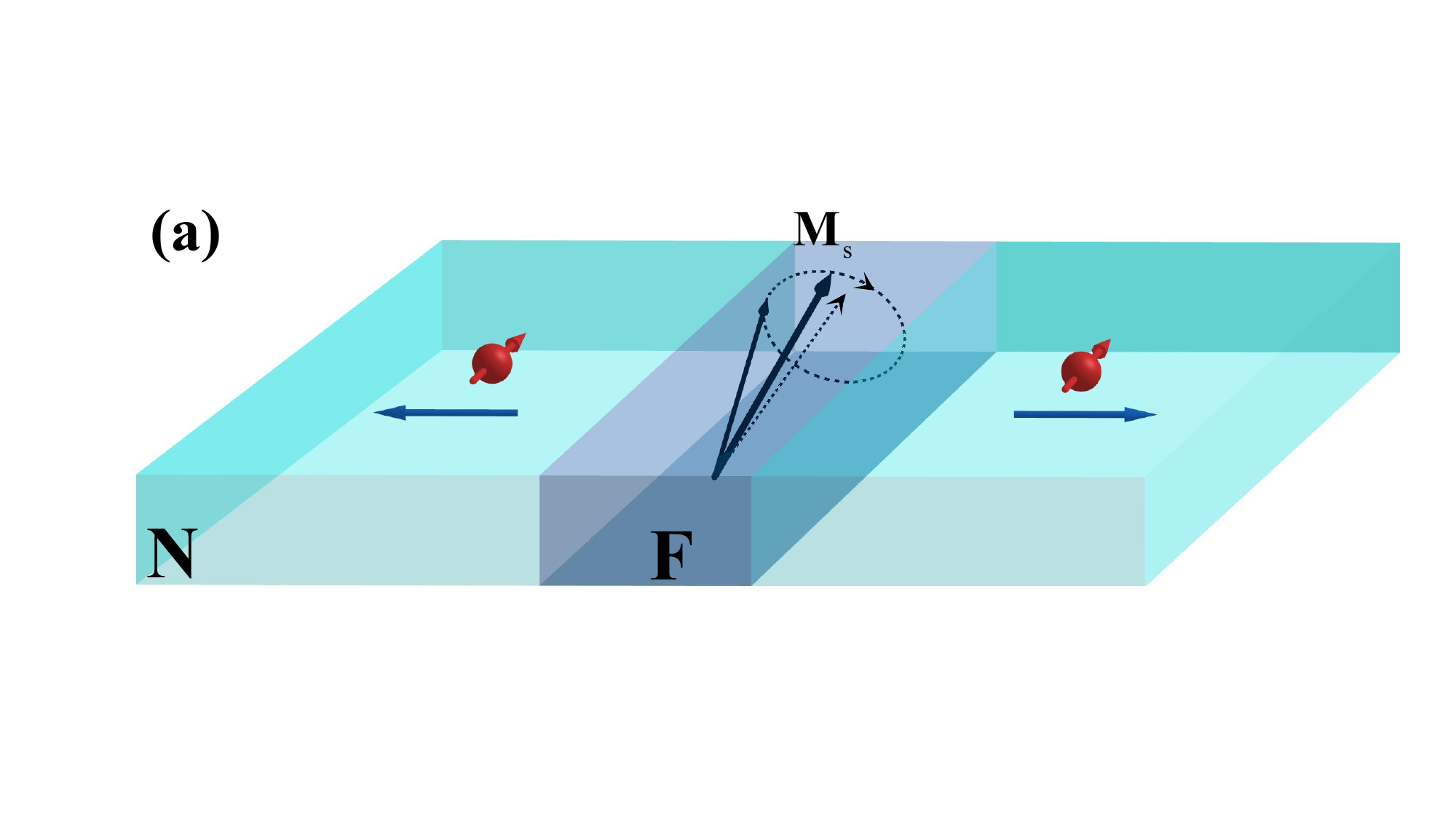}
    \vspace{5pt} 
    \includegraphics[width=0.96\linewidth, clip, trim=2cm 4cm 1cm 3cm]{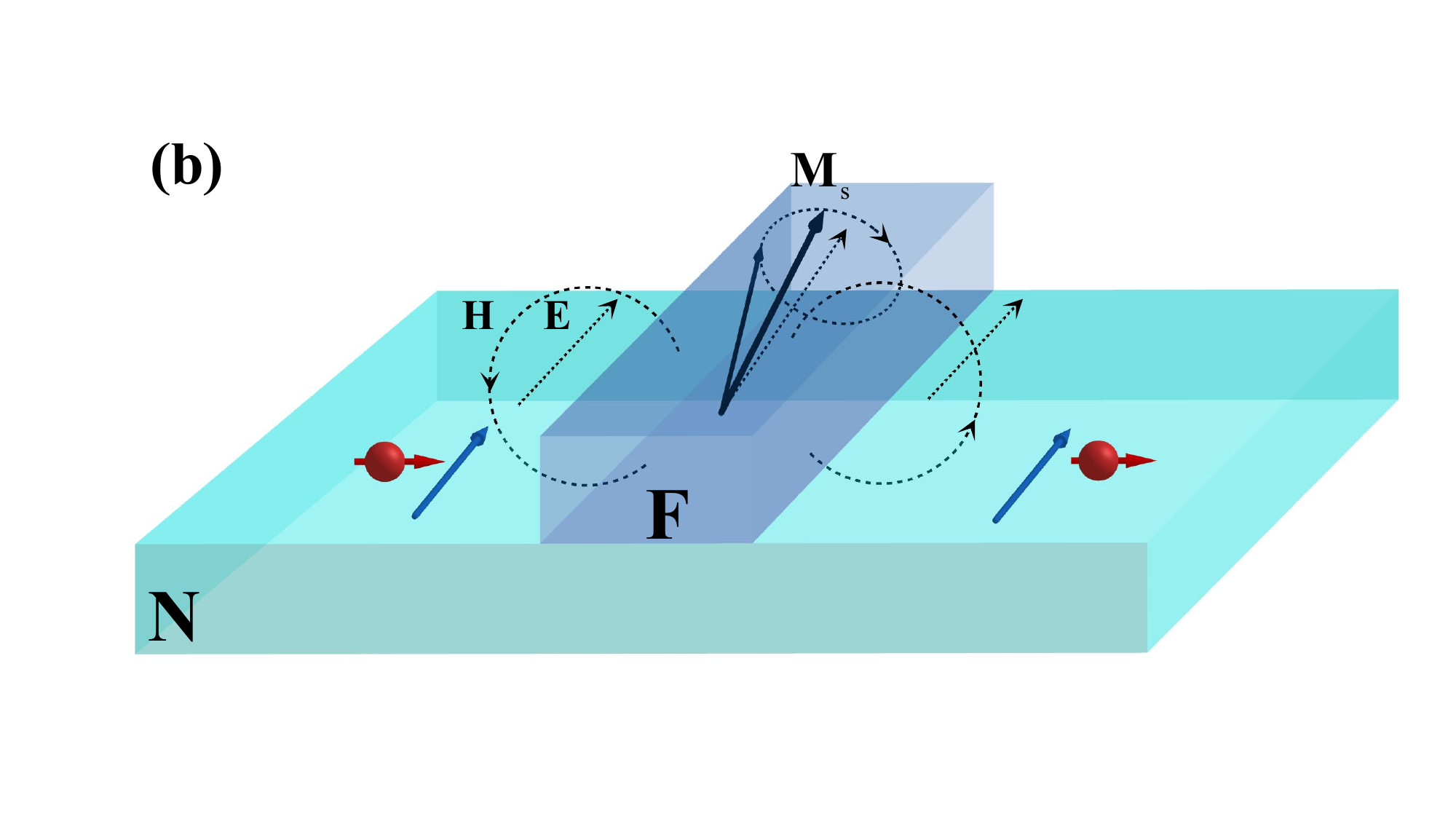}
    \caption{Comparison of longitudinal [(a)] and transverse or Hall-type [(b)] spin pumping by magnetization dynamics of the ferromagnet ``F" into the conductor ``N". ${\bf M}_s$ denotes the magnetization of the magnet. The blue and red arrows indicate, respectively, the flow and spin-polarization directions of a spin current. ${\bf H}$ and ${\bf E}$ in (b) denote the stray magnetic and electric fields emitted by the magnetization dynamics.}
    \label{Hall_spin_pumping}
\end{figure}

A central theme in magnetism and spintronics is to harness and manipulate the angular momentum current for applications such as the magnetization switching and information transmission~\cite{manchon2019current,chumak2015magnon,demidov2017magnetization,bauer2012spin,maekawa2023spin,sinova2015spin,bihlmayer2022rashba,ralph2008spin,chirality}. One efficient way to generate the flow of spin angular momentum or a spin current is the spin pumping~\cite{tserkovnyak2005nonlocal}, which, driven by the coherent magnetization dynamics, transfers the spin of localized electrons in the ferromagnets to itinerant electrons in conductors via the interfacial $s$-$d$ exchange interaction, inspiring significant advances in spin-based information processing and device engineering~\cite{tombros2007electronic,kajiwara2010transmission,yang2008giant,miron2011perpendicular,liu2012spin,caretta2018fast,grimaldi2020single,torrejon2017neuromorphic,blasing2020magnetic,luo2020current}. In a typical setup with the conductor$|$ferromagnet$|$conductor heterostructure, the magnetization dynamics pumps the spin current perpendicularly across the interfaces of the ferromagnet and conductor; the pumped spin current is the same in magnitude across the two interfaces of the ferromagnet and flows in opposite directions [refer to Fig.~\ref{Hall_spin_pumping}(a)], i.e., not unidirectional.

Besides localized electrons, many entities carry angular momentum, \textit{e.g.}, chiral phonons and circularly polarized electromagnetic fields. 
It was proposed or observed that the circular motion of the lattice can induce the spin polarization of electrons~\cite{chiral-phonon,KKim,XLi,NNishi} and orbital angular momentum current~\cite{orbital}. 
A focused time-varying magnetic field ${\bf H}$ can provide a Stern-Gerlach ``force" to pump a longitudinal spin current out of the source~\cite{spin-radiation}, with an efficiency governed by the photon spin ${\bf H}^*\times {\bf H}$. We call it and the conventional spin pumping~\cite{tserkovnyak2005nonlocal} as ``\textit{longitudinal spin pumping}". In such an intraband optical spin pumping~\cite{spin-radiation,Stefano_Optical}, the strong electric field in an electromagnetic field appears to play no direct role when the spin-orbit coupling (SOC) is weak (SOC may indirectly convey the optical orientation to a spin current by the interband optical transition~\cite{e-field1,e-field2,e-field3,e-field5,e-field7,e-field8,e-field9,e-field10,e-field11,e-field12,e-field13,e-field14,e-field15}), which strongly limits its efficiency.

In this work, we present an efficient paradigm for generating \textit{transverse} spin current via optical spin pumping, in which the electric and magnetic fields act jointly to significantly enhance the efficiency by several orders of magnitude over longitudinal optical spin pumping. To this end, we exploit the near-field electromagnetic radiation from magnetization dynamics in magnetic nanostructures (or focused laser beams). 
We find that, via the joint action of the electric and magnetic fields, the spin current flows parallel to the magnetic nanostructures and their interfaces with the conductor, i.e., a realization of transverse or Hall-type spin pumping. Figure~\ref{Hall_spin_pumping} compares the Hall-type spin pumping by the electromagnetic radiation with the longitudinal spin pumping by the interfacial exchange interaction. 
 In such electromagnetic radiation, the magnetic field also drives a longitudinal spin current. Both the transverse and longitudinal spin currents are unidirectional; the longitudinal current is locked to the magnetization direction (switching the magnetization reverses the spin current direction). The efficiency of the Hall-type/transverse spin pumping is several orders of magnitude larger than that of longitudinal spin pumping. This mechanism does not require photon angular momentum and provides an efficient pathway for generating spin currents for magnetization switching in low-dimensional spintronic devices.

This article is organized as follows. In Sec.~\ref{scattering_theory}, we develop a general scattering theory for the optical spin pumping and derive the pumped spin and charge currents by AC electromagnetic fields. We present the Hall-type/transverse and longitudinal spin pumpings in Sec.~\ref{transverse_spin_pumping} and \ref{longitudinal_spin_pumping}, respectively. An unidirectional charge current pumped by electromagnetic fields can arise due to the chirality of the electromagnetic field even in the absence of the SOC, which is addressed in Sec.~\ref{charge_pumping}. We summarize our results and give an outlook in Sec.~\ref{conclusion}.

\section{Scattering theory}
\label{scattering_theory}

A local electromagnetic field $\{{\bf E}({\bf r},t),{\bf H}({\bf r},t)\}$ can be generally considered as a time-dependent scattering potential for the motion of electrons. For a (quasi) two-dimensional electron gas (2DEG) that flows in the $x$-$y$ plane with a position vector ${\pmb \rho}=x\hat{\bf x}+y\hat{\bf y}$, the electric field ${\bf E}({\pmb \rho},t)=-\partial {\bf A}({\pmb \rho},t)/\partial t$ couples to the electron orbital motion via a vector potential ${\bf A}({\pmb \rho},t)$, while the magnetic field couples to the electron spin $\hat{\bf s}({\pmb \rho},t)$, governed by the Hamiltonian
\begin{align}
    \hat{H}=\frac{[\hat{\bf p}-e{\bf A}({\pmb \rho},t)]^2}{2m}+\mu_0\gamma_e{\bf H}({\pmb \rho},t)\cdot\hat{\bf s}({\pmb \rho}),
\end{align}
where $m$ is the effective mass of electrons, $\mu_0$ is the
vacuum permeability, and $\gamma_e$ is the gyromagnetic ratio of electrons. 
With the creation (annihilation) operator $\hat{c}^\dagger_{{\bf k}\alpha}$ ($\hat{c}_{{\bf k}\alpha}$) of electrons with wave vector ${\bf k}$ and spin $\alpha=\{\uparrow,\downarrow\}$ along $\hat{\bf z}$, the free Hamiltonian $\hat{H}_0=\sum_{{\bf k}}\sum_{\alpha=\uparrow,\downarrow}\left[{\hbar^2k^2}/({2m})-\mu\right]\hat{c}^\dagger_{{\bf k}\alpha}\hat{c}_{{\bf k}\alpha}$
is described by a spin-degenerate parabolic band with respect to the chemical potential $\mu$. The electromagnetic spin pumping proposed here does not rely on the SOC, which can be straightforwardly included in the present formalism.   
The monochromatic electromagnetic field of frequency $\omega$, i.e., ${\bf E}({\pmb \rho},t)=\sum_{\bf q}({\bf E}^+({\bf q})e^{-i\omega t}+{\bf E}^-({\bf q})e^{i\omega t})e^{i{\bf q}\cdot {\pmb \rho}}$ and  ${\bf H}({\pmb \rho},t)=\sum_{\bf q}({\bf H}^+({\bf q})e^{-i\omega t}+{\bf H}^-({\bf q})e^{i\omega t})e^{i{\bf q}\cdot {\pmb \rho}}$, mediates a coupling between electrons of different wave vectors:
\begin{align}
    {\hat V}(t)=\sum_{{\bf k}{\bf k}^\prime}\sum_{\xi=\pm}\sum_{\alpha\beta}\mathcal{G}^\xi_{\alpha\beta}({\bf k},{\bf k}^\prime)\hat{c}^\dagger_{{\bf k}\alpha}\hat{c}_{{\bf k}^\prime\beta}e^{-i\xi\omega t},
\end{align}
in which the coupling constant matrix for the crystal area $A$ in the spin space 
\[
\mathcal{G}^\xi({\bf k},{\bf k}^\prime)=\frac{i\xi e\hbar}{2m\omega A}({\bf k}+{\bf k}^\prime)\cdot{\bf E}^\xi({\bf k}-{\bf k}^\prime)+\frac{\mu_0\gamma_e\hbar}{2A}{\bf H}^\xi({\bf k}-{\bf k}^\prime)\cdot{\pmb\sigma},
\] 
in which $\xi=``+"$ and $``-"$ correspond to the photon emission and absorption processes.

Treating the local electromagnetic field as a local scattering potential, the incident electron $\hat{c}_{{\bf k^\prime}\alpha}$ of an eigenstate of $\hat{H}_0$ is scattered to the scattering state $\hat{b}_{{\bf k}\beta}=\sum_{{\bf k}^\prime\alpha}T_{\beta\alpha}({\bf k},{\bf k}^\prime,t)\hat{c}_{{\bf k}^\prime\alpha}$ according to the $T$-matrix 
 \begin{align}
 \hspace{-0.15cm}&T_{\beta\alpha}({\bf k}^\prime,{\bf k},t)=\delta_{{\bf k}^\prime{\bf k}}\delta_{\beta\alpha}+\sum_{\xi=\pm} \Gamma_{\beta\alpha}^{\xi}({\bf k}^\prime,{\bf k})e^{-i(\xi\hbar\omega-\varepsilon_{k^\prime}+\varepsilon_k)t/\hbar}\nonumber\\
  \hspace{-0.15cm}&+\sum_{\xi_1,\xi_2=\pm}\Delta_{\beta\alpha}^{\xi_1\xi_2}({\bf k}^\prime,{\bf k})e^{-i(\varepsilon_k+(\xi_1+\xi_2)\hbar\omega-\varepsilon_{k^\prime})t/\hbar},
\end{align}
obtained by the evolution of the electron wavefunction under the interaction $\hat{V}(t)$. 
Interpreted by the scattering theory,  
\begin{align}
    \Gamma_{\beta\alpha}^\xi({\bf k}^\prime,{\bf k})=\frac{\mathcal{G}^\xi_{\beta\alpha}({\bf k}^\prime,{\bf k})}{\xi\hbar\omega-\varepsilon_{k^\prime}+\varepsilon_k+i\delta}
\end{align}
denote the scattering amplitudes when the electron emits ($\xi=+$) or absorbs ($\xi=-$) one photon, in which $\delta\rightarrow0^+$ is introduced due to the adiabatic introduction of the interaction, while the amplitudes  
\begin{align}
    \Delta_{\beta\alpha}^{\xi_1\xi_2}({\bf k}^\prime,{\bf k})
    &=\sum_{{\bf q}}\sum_{\gamma}\frac{\mathcal{G}^{\xi_1}_{\beta\gamma}({\bf k}^\prime,{\bf q})}{\left(\varepsilon_{q}-\varepsilon_k-\xi_2\hbar\omega-i\delta\right)}\nonumber\\
    &\times\frac{\mathcal{G}^{\xi_2}_{\gamma\alpha}({\bf q},{\bf k})}{ \left(\varepsilon_{k^\prime}-\varepsilon_k-(\xi_1+\xi_2)\hbar\omega-i\delta\right)}
\end{align}
are the scattering amplitudes involving the two-photon processes. With the operator $\hat{\pmb b}_{{\bf k}}(t)=(\hat{b}_{{\bf k}\uparrow}(t),\hat{b}_{{\bf k}\downarrow}(t))^T$, the field operator of electron evolves according to $\hat{\Psi}({\pmb \rho},t)
    =({1}/{\sqrt{A}})\sum_{\bf k}(\chi_\uparrow,\chi_\downarrow)e^{i{\bf k}\cdot{\pmb\rho}}\hat{\pmb b}_{\bf k}(t)$,
where $\chi_{\uparrow,\downarrow}$ is the spinor wavefunction of electrons.

Substitution of the field operator into the spin-current operator
\begin{align}
\hat{\pmb {\cal J}}_s({\pmb\rho},t)={\hbar^2}/({4im})\hat{\Psi}^\dagger{\pmb\sigma}\otimes\nabla\hat{\Psi}+\rm{H.c.}
\end{align}
    and charge-current operator 
\begin{align}
    \hat{\bf J}_e({\pmb\rho},t)={e\hbar}/({2im})\hat{\Psi}^\dagger\nabla\hat{\Psi}+\rm{H.c.}
\end{align}
and performing the ensemble average $\langle\hat{c}^\dagger_{{\bf k}\alpha}\hat{c}_{{\bf k}^\prime\beta}\rangle=\delta_{{\bf k}{\bf k}^\prime}\delta_{\alpha\beta}f(\varepsilon_k)$ in terms of the Fermi-Dirac distribution $f(\varepsilon_k)$
yields the equilibrium, AC, and DC spin/charge currents pumped by the AC electromagnetic field.

The DC spin current is useful for driving the dynamics of magnetization~\cite{tombros2007electronic,kajiwara2010transmission,yang2008giant,miron2011perpendicular,liu2012spin,caretta2018fast,grimaldi2020single,torrejon2017neuromorphic,blasing2020magnetic,luo2020current}, which reads
\begin{align}
&{\pmb {\cal J}}_{s}(\boldsymbol{\rho}) 
=\frac{i\hbar^4\mu_0\gamma_e}{8mA^3}\sum_{{\bf k}{\bf k}^\prime{\bf q}}\sum_{\xi=\pm} e^{i({\bf k}^\prime-{\bf k})\cdot\boldsymbol{\rho}}\nonumber\\
&\times\Big[\mu_0\gamma_e\left({\bf H}^{-\xi}({\bf k}'-{\bf q})\times{\bf H}^{\xi}({\bf q}-{\bf k})\right)\otimes{\bf k}^\prime \nonumber\\
&-\frac{\xi e}{m\omega}\left(({\bf k}'+{\bf q})\cdot{\bf E}^{-\xi}({\bf k}'-{\bf q})\right){\bf H}^{\xi}({\bf q}-{\bf k}) \otimes({\bf k}+{\bf k}')\Big]\nonumber\\
&\times \frac{ f(\varepsilon_{q})-f(\varepsilon_{k})}{\left(\varepsilon_{ k'}-\varepsilon_{ q}+\xi\hbar\omega-i\delta\right)\left(\varepsilon_{ k}-\varepsilon_{ q}+\xi\hbar\omega+i\delta\right)}+{\rm H.c.}.
\label{nonlinear_spin_current}
\end{align}
The electric field itself does not contribute to the spin current in the absence of SOC, since it conserves electron spin.  
The local magnetic field pumps the longitudinal spin current ${\pmb  {\cal J}}^L_s$ that is polarized along ${\bf H}^{-\xi}\times{\bf H}^\xi\sim {\bf H}^{\xi*}\times{\bf H}^\xi$ and flows along the gradient of the magnetic field, which may be interpreted by the transfer of photon spin~\cite{photon_spin} to electron spin current, as predicted before~\cite{spin-radiation}. 
Surprisingly, even though it does not correspond to a photon spin, the joint effect of the electric and magnetic fields also leads to a pure DC spin current, which is much more efficient because electrons couple strongly to the electric field. In this scenario, the spin polarization is aligned with the magnetic field, while the spin current flows in the direction of the electric field. In a local transverse electromagnetic field, the electric field is transverse to the field gradient; thus, such a spin current is a transverse or Hall response that flows perpendicular to the field gradient. Its existence depends on the dimension of the electromagnetic field: it generally occurs in two dimensions, whereas in one-dimensional fields the current arises when a phase difference different from $\pi$ exists between the electric and magnetic components (refer to Appendix~\ref{appendix_conditions} for a detailed discussion).

The existence of the Hall-type/transverse spin pumping can be intuitively understood as follows. By definition, in the configuration shown in Fig.~\ref{Hall_spin_pumping}(b)  the \textit{transverse} spin current density $\propto {\bf v}_y({\pmb \rho},t)\otimes{\bf s}^\nu({\pmb \rho},t)$, in which $\hat{\bf y}$ is parallel to the magnetic nanostructures and $\nu=\{x,y,z\}$ is the spin indices. The electric field drives the velocity density ${\bf v}({\pmb \rho},t)$ along the electric field, while the magnetic field polarizes the spin density ${\bf s}^\nu({\pmb \rho},t)$ along the magnetic field. Although both the velocity and spin densities are time-dependent, their time averages over the AC field period may exist, yielding a transverse DC spin current.

Our predicted transverse and longitudinal spin pumping by electromagnetic fields differ from the inverse Faraday effect~\cite{Pershan,Hertel} and optical spin-transfer torque~\cite{Rossier,Nvemec}. Both the inverse Faraday effect and optical spin-transfer torque are solely governed by the electric field, while our predicted transverse spin pumping $\sim {\rm Im}({\bf E}^*\otimes {\bf H})$ and unidirectional longitudinal spin pumping $\sim ({\bf H}^*\times {\bf H})$ by the evanencent electromagnetic fields involves the magnetic component. Moreover, both the inverse Faraday effect and optical spin-transfer torque focus on the static magnetization generated by the optical field rather than the spin current. Indeed, in the inverse Faraday effect driven by circularly polarized light, the electrons exhibit oscillating electron velocity and oscillating electron density, generating a static magnetic moment $\propto {\bf E}\times{\bf E}^*$~\cite{Pershan,Hertel}. In the optical spin-transfer torque, the circularly polarized laser pulses excite net spin angular momentum carried by the electron-hole pairs across the band gap of the ferromagnetic semiconductor [such as (Ga, Mn)As] via the optical selection rules, which then interact with the magnetic moments via the exchange coupling and lead to a torque on the magnetization~\cite{Rossier,Nvemec}.

The time-varying electromagnetic field may rectify a DC charge current, which is an important observable in experiments~\cite{canming_hu}. It reads
\begin{align}
&{\bf J}_{e}({\pmb\rho})    
=\frac{e\hbar^3}{2mA^3}\sum_{{\bf k}{\bf k}^\prime{\bf q}}\sum_{\xi=\pm}{\bf k}^\prime e^{i({\bf k}^\prime-{\bf k})\cdot{\pmb\rho}}\nonumber\\ 
&\times\bigg{[}\frac{\mu_0^2\gamma_e^2}{2}{\bf H}^{-\xi}({\bf k}^\prime-{\bf q})\cdot{\bf H}^\xi({\bf q}-{\bf k})\nonumber\\
&+\left(\frac{e}{2m\omega}\right)^2({\bf k}^\prime+{\bf q})\cdot{\bf E}^{-\xi}({\bf k}^\prime-{\bf q})({\bf q}+{\bf k})\cdot{\bf E}^\xi({\bf q}-{\bf k})\bigg{]}\nonumber\\
&\times\frac{f(\varepsilon_{k})-f(\varepsilon_{q})}{\left(\varepsilon_{q}-\varepsilon_{k^\prime}-\xi\hbar\omega+i\delta\right)\left(\varepsilon_{ k}-\varepsilon_{ q}+\xi\hbar\omega+i\delta\right)}+{\rm H.c.}.
\label{nonlinear_charge_current}
\end{align}
Such a DC charge current arises exclusively from quadratic contributions of pure electric or purely magnetic fields, with the flow direction governed by the electromagnetic field gradients, indicating that the transverse spin current in $\pmb{\cal J}_s$ is ``pure".

\section{Hall-type spin pumping}

\label{transverse_spin_pumping}

We substantiate the expectation of pure transverse spin current resulting from the joint effect of electric and magnetic fields by experimentally convenient realizations.

\begin{figure}[htp!]
    \centering
    \includegraphics[width=0.485\linewidth]{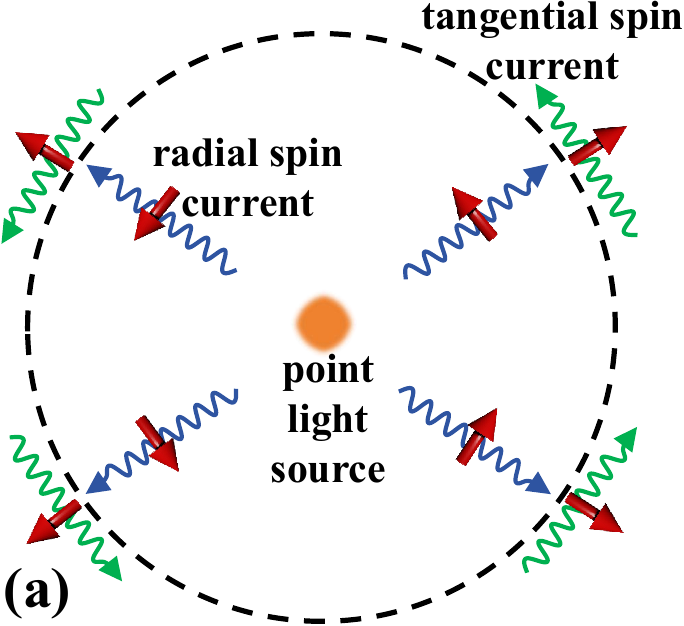}
    \includegraphics[width=0.485\linewidth]{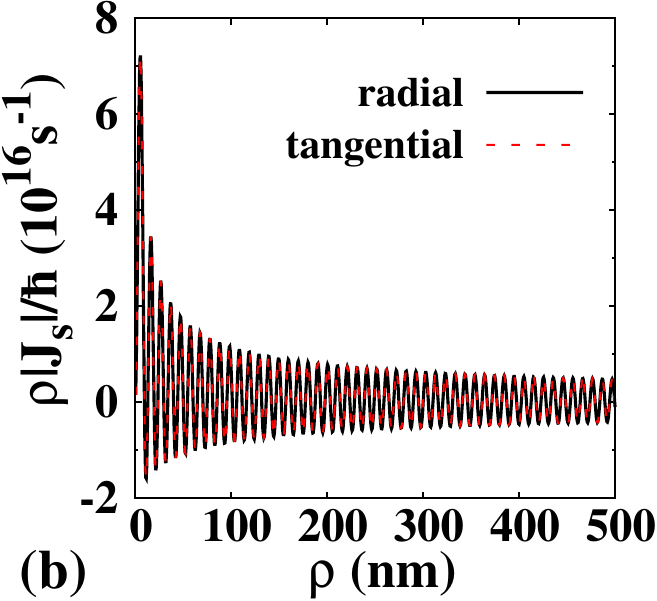}
    \caption{Spin current pumped by the joint effect of the electric and magnetic fields of the point light source in the 2DEG. (a) illustrates the flow and the spin-polarization directions. (b) shows the oscillation of $\rho|{\pmb{\cal J}}_{s}(\pmb{\rho})|$ away from the source.}
    \label{Driven_Js}
\end{figure}

The electric ${\bf E}({\boldsymbol{\rho}},t)\simeq E_0\delta(\rho)e^{-i\omega t}\left(1,i,0\right)^T+{\rm H.c.}$ and magnetic field ${\bf H}({\boldsymbol{\rho}},t)\simeq{E_0}/({\mu_0c})\delta(\rho)e^{-i\omega t}\left(i,-1,0\right)^T+{\rm H.c.}$ in a Gaussian optical beam of small radius, where $E_0$ is the amplitude of the electric field, jointly drives the spin current in the 2DEG (refer to Appendix~\ref{appendix_optical_pumping} for a derivation of the spin current) 
    \begin{align}
    {\pmb {\cal J}}_s(\pmb{\rho})
    &=\frac{me\gamma_eE_0^2q_F}{64\hbar\pi c\rho}\left((\hat{\bf e}_\perp\otimes\hat{\bf e}_\rho)+(\hat{\bf e}_\rho\otimes\hat{\bf e}_\perp)\right)\nonumber\\
    &\times\left(J_{0}(q_F\rho)H_{1}(q_F\rho)-J_{1}(q_F\rho)H_{0}(q_F\rho)\right),
\end{align}
where $J_{n}(x)$ and $H_{n}(x)$ are the $n$-order Bessel function of the first kind and Struve function and $q_F$ is the Fermi wave vector of the 2DEG. 
As in Fig.~\ref{Driven_Js}(a), the driven radially flowing spin current carries a tangential spin polarization, while the tangentially flowing spin current carries a radial spin polarization. These two types of spin current show the same dependence on distance away from the source, oscillating with the period $\lambda_F=2\pi/q_F$, as indicated by $\rho|{\pmb{\cal J}}_{s}(\pmb{\rho})|$ in Fig.~\ref{Driven_Js}(b). Cruz and Flatt\'e predicted a circulating spin current with oscillations induced by a spin defect, relying on a different mechanism based on the SOC~\cite{Adonai}. In the calculation, we use the parameters of  $n$-doped $\mathrm{MoS_{2}}$ with effective mass of electron 
$m=0.48m_{e}$~\cite{MoS_2_mass}, $g$-factor $|g_{e}
|=2.16$~\cite{MoS_2_g}, and the chemical potential $\mu=30$~meV, noting there are two equivalent valleys carrying spin current. The field frequency $\omega=10$~THz and its amplitude $E_{0}=\pi\times10^{-9}$~V/m is equivalent to a spot of laser beams of radius $100$~nm with electric field $1$~kV/cm. Such a chosen electric field with amplitude $E_0=1~{\rm kV/cm}$ is easily achievable by the local fields generated by a focused THz optical beams since the peak amplitudes in the range of $10\sim100~{\rm kV/cm}$ were routinely reported~\cite{electric_amplitude1,electric_amplitude2}. 

We find that the fast oscillation of the spin current vanishes when it is pumped by a one-dimensional electromagnetic field. Such an electromagnetic field can be conveniently generated by the ferromagnetic resonance (FMR)~\cite{EP_effect,Y_Au} of a magnetic nanowire of thickness $d$ and width $w$, biased by the magnetic field ${\bf H}_0$ along the wire $\hat{\bf y}$-direction, as illustrated in Fig.~\ref{model}.
Under a resonant excitation,
the magnetization ${\bf M}=(i\zeta^2M_z,0,M_z)e^{-i\omega_{\rm K} t}$ precesses with the Kittel frequency $\omega_{\rm K}$ and ellipticity $\zeta^2$, depending on the wire geometry.
The magnetization dynamics generates the magnetization current ${\bf J}_M=\nabla\times {\bf M}$ that radiates the electric field~\cite{Jackson} 
\begin{align}
    \mathbf{E}(\mathbf{r},t)=\frac{i \mu_0 \omega}{4 \pi} \int \frac{\left[\nabla^{\prime} \times \mathbf{M}\left(\mathbf{r}^{\prime},t\right)\right] e^{i k\left|\mathbf{r}-\mathbf{r}^{\prime}\right|}}{\left|\mathbf{r}-\mathbf{r}^{\prime}\right|} d \mathbf{r}^{\prime},\label{stray-electric-field}
\end{align}
where $k=\omega/c$ is the wave number of microwaves. In the 2DEG $(z=0)$, the Fourier components 
\begin{align}
   {\bf E}^{\xi}(k_x,k_y)
    &=-i\frac{\mu_0\omega}{2} (1-e^{-\left|k_x\right| d})\frac{ \sin \left(k_x w / 2\right)}{k_x^2}\nonumber\\
    &\times\left(1-\xi\zeta^2{\rm sgn}(k_x) \right)M_z\delta(k_y)\hat{\bf y}
    \label{electric_field_k}
\end{align} in $\mathbf{E}({\bf r},t)|_{z=0}=\sum_{\xi=\pm}\sum_{k_x}{\bf E}^{\xi}(k_x,k_y)e^{i(k_xx+k_yy-\xi\omega t)}$
are along the wire $\hat{\bf y}$-direction and strongly depend on the ellipticity, i.e., they propagate \textit{unidirectionally} along the negative $\hat{\bf x}$-direction when $\zeta^2=1$ ($d=w$), but propagate bidirectionally when $\zeta^2\rightarrow 0$ ($d\ll w$).
The stray magnetic field is emitted by the FMR via $\nabla \times {\bf E}=-\mu_0{\partial {\bf H}}/{\partial t}$; 
\begin{align}
&\left(\begin{array}{cc}
H^{\xi}_z(k_x,k_y)\\
H^{\xi}_x(k_x,k_y)\end{array}\right)
=-\xi\frac{i}{2} (1-e^{-\left|k_x\right| d})\frac{ \sin \left(k_x w / 2\right)}{k_x}\nonumber\\
&\times\left(1-\xi\zeta^2{\rm sgn}(k_x) \right)\left(\begin{array}{cc}1\\
i{\rm sgn}(k_x)\end{array}\right)M_z\delta(k_y)
 \label{manegtic_field_k}
\end{align}
are circularly polarized and normal to the wire $\hat{\bf y}$-direction.

\begin{figure}[htp!]
    \centering
    \includegraphics[width=0.95\linewidth, clip, trim=0cm 0cm 0cm 0cm]{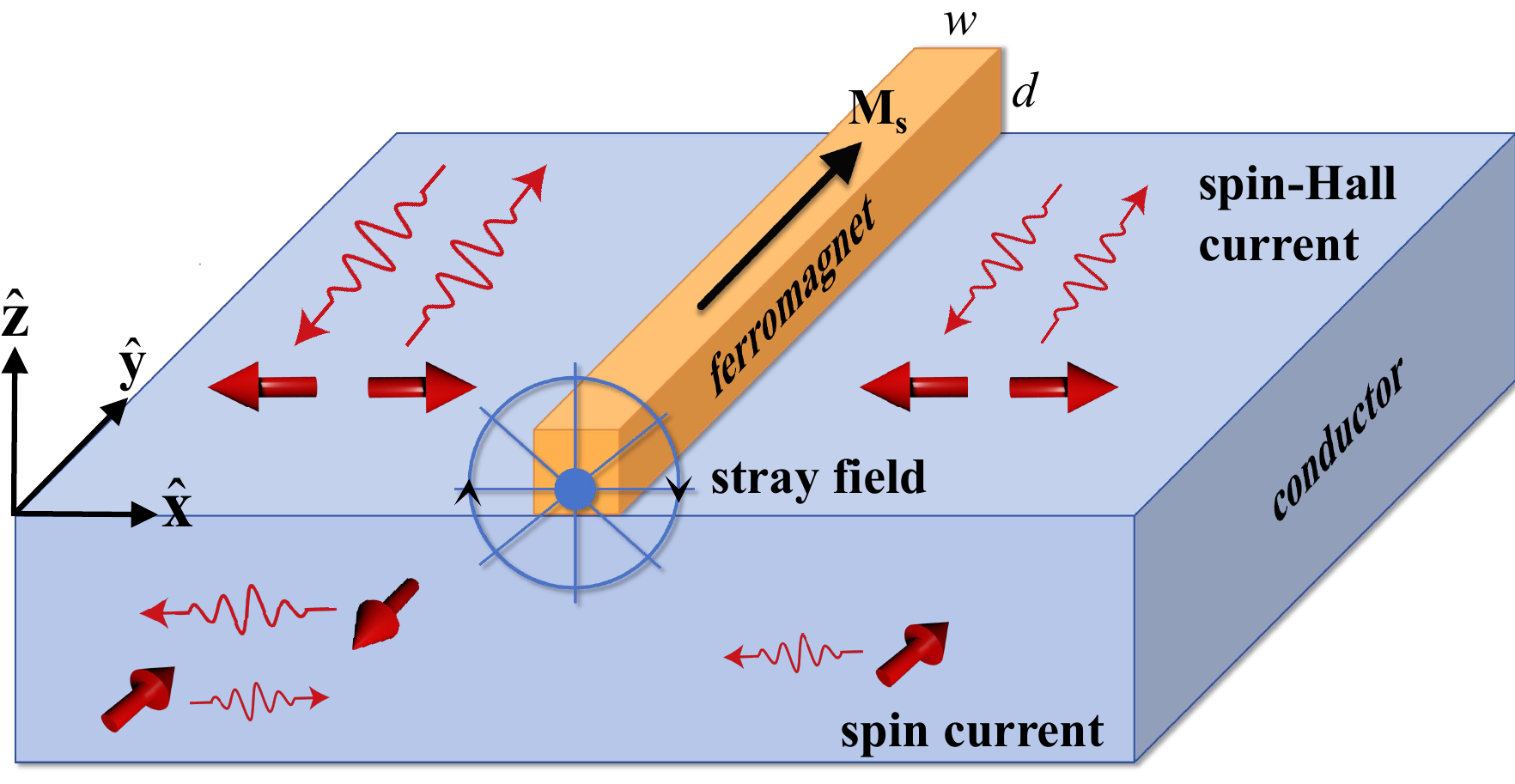}   
    \label{model}
    \caption{Pumping of transverse and longitudinal spin currents by the stray electromagnetic field. The thick red arrows indicate the spin-polarization direction, and the red curves denote the direction of electron propagation. Both the transverse and longitudinal spin currents are unidirectional. }  
    \label{model}
\end{figure}

\begin{figure}[htp]
    \centering
    \includegraphics[width=0.485\linewidth, clip, trim=0cm 0cm 0cm 0cm]{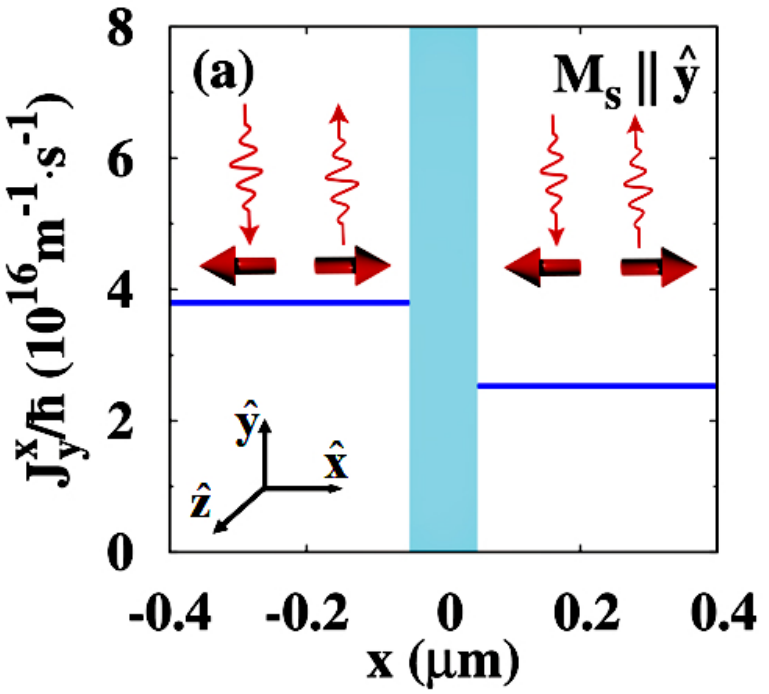}
    \includegraphics[width=0.485\linewidth, clip, trim=0cm 0cm 0cm 0cm]{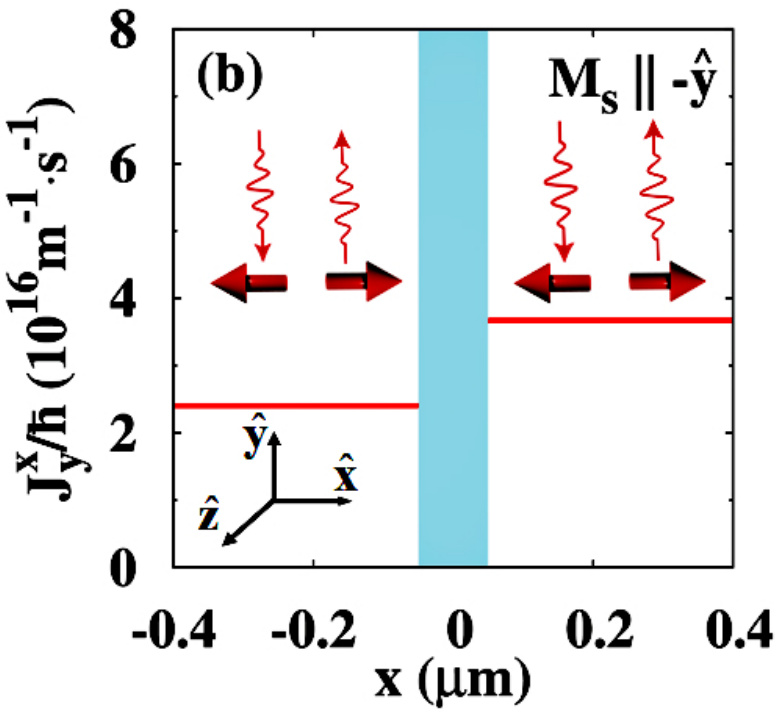}
     \includegraphics[width=0.49\linewidth, clip, trim=2cm 0cm 2.8cm 1cm]{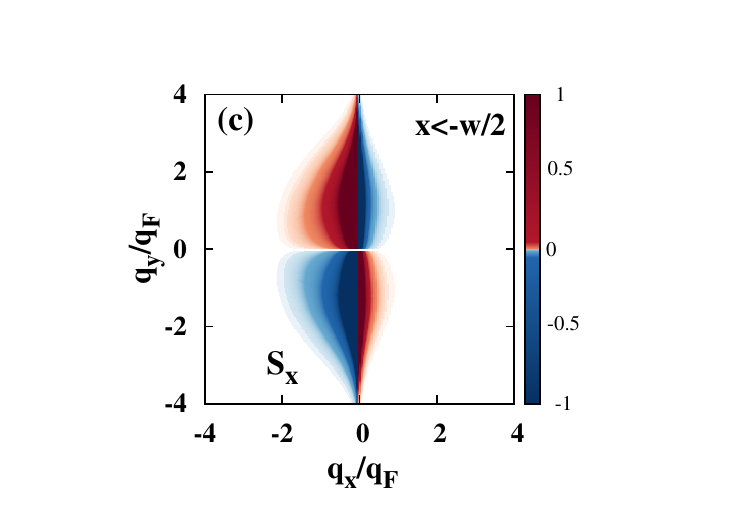}
     \includegraphics[width=0.49\linewidth, clip, trim=2cm 0cm 2.8cm 1cm]{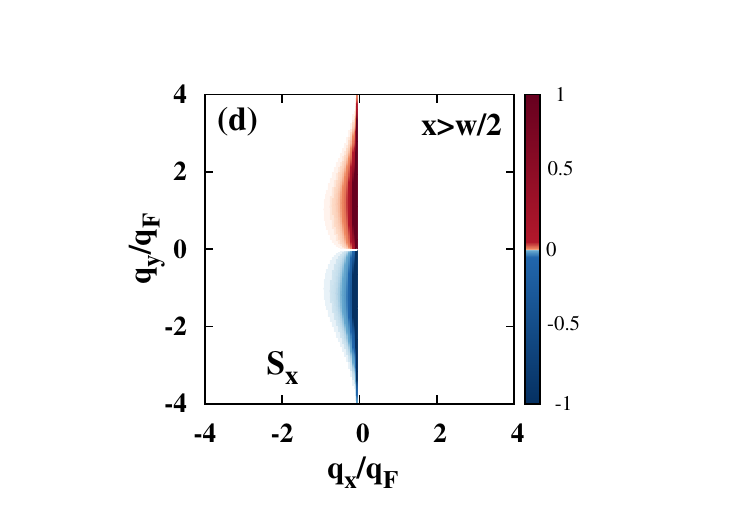}
    \caption{Transverse unidirectional spin current pumped by magnetization dynamics. (a) and (b)  illustrate the spatial distribution of  Hall spin current density pumped by a near electromagnetic field when ${\bf M}_s\parallel\hat{\bf y}$ and ${\bf M}_s\parallel -\hat{\bf y}$, respectively. The blue rectangular area indicates the region covered by the nanowire. (c) and (d) depict the asymmetric spin texture ${\bf S}_x(q_x,q_y)$ in the wave-vector space (normalized by its maximum value) that differ at $x<-w/2$ and $x>w/2$, demonstrating the physical origin of the Hall spin current with $\sum_{q_x}{\bf S}_x(q_x,q_y)>0$ [$\sum_{q_x}{\bf S}_x(q_x,q_y)<0$] for $q_y>0$ ($q_y<0$). Parameters used for calculation are given in the text.}
    \label{Spinhall}
\end{figure}

Figure~\ref{Spinhall}(a) and (b) illustrate the Hall-type spin pumping in a 2DEG (e.g., ${{\rm Mo S}_2}$ using parameters in Fig.~\ref{Driven_Js}) pumped by the FMR $\omega_{\rm K}=2\pi\times28.15 $~GHz of CoFeB nanowire of $d=w=100~{\rm nm}$. The saturation magnetization $\mu_0M_s=1.6{\rm\ T}$~\cite{CoFeB_1,CoFeB_2,CoFeB_3} is biased by a static magnetic field $\mu_0H_0=0.1\ {\rm T}$ along the wire $\pm \hat{\bf y}\mbox{-}$direction, and a small transverse magnetization $M_z=0.1M_s$ is excited by uniform microwaves. Such a precession amplitude $M_z=0.1 M_s$ corresponds to the cone angle  $\theta\approx5.7^\circ$ by  $M_z/M_s=\sin\theta=0.1$, which lies well within the reported FMR experiments 
since cone angles in the range of $6^\circ\sim 9^\circ$ have been well reported in experimental studies of various nanostructured ferromagnet~\cite{cone_angle1,cone_angle2,cone_angle3,cone_angle4}. 
As illustrated in Fig.~\ref{Spinhall}(a), the flow direction of electrons is locked to the direction of spin polarization, that is, the electrons flowing along the $+\hat{\bf y}$/$-\hat{\bf y}$-direction carry $+{\hat{\bf x}}$/$-{{\hat{\bf x}}}$ spin polarization, which results in a pure spin current since the net charge current vanishes according to Eq.~\eqref{nonlinear_charge_current}. Although the magnetic field has both the $x$ and $z$ components [Eq.~\eqref{manegtic_field_k}], only the $x$-component polarizes the spin since it holds a $\pi/2$-phase difference relative to the electric field.

The generation of Hall-type spin current can be well understood by analyzing the spin texture ${\bf S}(q_x,q_y)$ of electrons pumped by the electromagnetic field, an intuition inspired by the spin-Hall effect~\cite{spin_hall_effect_RSOC}. To this end, we decompose the transverse spin current in Eq.~\eqref{nonlinear_spin_current} according to $\pmb{\cal J}^{T}_s(x)=\sum_{q_x,q_y}{\bf v}(q_x,q_y)\otimes {\bf S}(q_x,q_y,x)$, in which ${\bf v}(q_x,q_y)$ is the group velocity of electrons and ${\bf S}(q_x,q_y,x)$ is the spin texture we want to define. By definition, we need to analyze this spin texture by distinguishing the position $x<-w/2$ and $x>w/2$. Figure~\ref{Spinhall}(c) and (d) illustrate the spin texture in momentum space at the left- and right-hand side of the magnetic nanowire. Only the $x$-component ${\bf S}_x$ exists, demonstrating the excited spin current is polarized along $\hat{\bf x}$. Furthermore, for electrons moving in the $+\hat{\bf y}$ direction $(q_y>0)$, the net spin polarization $\sum_{q_x}{\bf S}_x(q_x,q_y)$ is oriented along $+\hat{\bf x}$, while those moving in the $-\hat{\bf y}$ direction carry a net $-\hat{\bf x}$ polarization, explaining exactly the properties of the pumped Hall spin current.

Another way to intuitively understand the Hall spin pumping is to analyze the excited velocity and spin densities by electromagnetic fields. Under the local electric field ${\bf E}({\bf r},t)$ of the frequency $\omega$ that scatters the electrons between the wave vectors ${\bf k}$ and ${\bf k}'$ with the scattering amplitude  $\mathcal{G}^\xi({\bf k}^\prime,{\bf k})$, the electron velocity $\hat{\bf v}({\pmb \rho},t)=\hbar/(2im)\hat{\Psi}^\dagger{\pmb\nabla}\hat{\Psi}+{\rm H.c.}$ in the linear response regime is found to be 
\begin{align}
    {\bf v}({\boldsymbol{\rho},t})&=\frac{\hbar}{2mA}\sum_{{\bf k}{\bf k}^\prime}\sum_{\xi=\pm}\sum_{\alpha,\beta=\uparrow\downarrow}({\bf k}+{\bf  k}^\prime)  \mathcal{G}^\xi_{\alpha\beta}({\bf k}^\prime,{\bf k})\nonumber\\
    &\times\frac{e^{i\left(({\bf k}^\prime-{\bf k})\cdot {\pmb \rho}-\xi\omega t)\right)}}{\xi\hbar\omega-\varepsilon_{ k^\prime}+\varepsilon_{ k} +i\delta}
     f(\varepsilon_k)+{\rm H.c.}.
\end{align}
In the configuration of Fig.~\ref{model}, the electric field is along the magnetic nanowire $\hat{\bf y}\mbox{-}$direction with Fourier components 
${\bf E}^\xi({\bf k})=2\pi E^\xi(k_x)\delta(k_y)\hat{\bf y}$, so the driven electron velocity is along the wire direction with  
\begin{align}
{\bf v}_y(x,t)
&=\frac{e}{m\omega A}\sum_{{k_x} k_y}\sum_{\xi=\pm}\xi\frac{k_y^2}{k_\xi}E^\xi(k_\xi-k_x)\nonumber\\
&\times e^{i\left[({ k}_\xi-{ k_x})x-\xi\omega t)\right]}f(\varepsilon_{k})\hat{\bf y}+{\rm H.c.},
\label{velocity}
\end{align}
where $k_{\xi}=\sqrt{k_x-2\xi m\omega/\hbar}$.
On the other hand, the magnetic field ${\bf H}({\bf r},t)$ excites the electron spin density $\hat{\bf  s}({\boldsymbol{\rho}},t)=({\hbar}/2)\hat{\Psi}^\dagger{\pmb\sigma}\hat{\Psi}$ in the linear response regime according to 
\begin{align}
    {\bf s}^\nu({\boldsymbol{\rho}},t)&=\frac{\mu_0\hbar^2\gamma_e}{4 A^2}\sum_{{\bf q}{\bf q}^\prime}\sum_{\xi=\pm}\sum_{\alpha,\beta=\uparrow\downarrow}{{\sigma}}^\nu_{\alpha\beta}[{\pmb\sigma}_{\beta\alpha}\cdot  {\bf H}^\xi({\bf q}^\prime-{\bf q})]\nonumber\\
    &\times\frac{e^{i\left[({\bf q}^\prime-{\bf q})\cdot{\pmb\rho}-\xi\omega t)\right]}}{\xi\hbar\omega-\varepsilon_{q^\prime}+\varepsilon_{q} +i\delta}
     f(\varepsilon_{q}){\hat{\pmb\nu}}+{\rm H.c.}.
\end{align}
For the configuration in Fig.~\ref{model}, the excited spin density 
\begin{align}
    {\bf s}^\nu(x,t)&=-\frac{i\mu_0m\gamma_e}{2 A}\sum_{q_xq_y}\sum_{\xi=\pm}\frac{1}{q_\xi} H_\nu^\xi(q_\xi-q_x)\nonumber\\
    &\times e^{i\left[({ q}_\xi-{ q_x})x-\xi\omega t)\right]}
     f(\varepsilon_{q}){\hat{\pmb\nu}}+{\rm H.c.}.
     \label{excited spin}
\end{align}
Accordingly, the driven electron velocity and spin density evolve according to 
\begin{align}
    {\bf v}_y(x,t)\propto&\xi E^\xi[\sin((k_\xi-k_x)x)\nonumber\\
    +&i\cos((k_\xi-k_x)x)]e^{-i\xi\omega t}+{\rm H.c.},\nonumber\\
    {\bf s}^\nu(x,t)\propto& i H_\nu^\xi[\sin((q_\xi-q_x)x)\nonumber\\
    +&i\cos((q_\xi-q_x)x)]e^{-i\xi\omega t}+{\rm H.c.}.
\end{align}
For the observable real part of ${\bf v}_y\otimes{\bf s}^\nu$, the DC transverse spin current $\sim {\rm Re}(i\xi E^{-\xi} H^{\xi}_\nu)$, which is thereby governed by the relative phase between $E^{-\xi}$ and $H^\xi$. This explains that the transverse spin current arises only when a phase difference exists between the electric and magnetic components in the configuration shown in Fig.~\ref{model}.

The flow direction of the Hall spin current is the same at the two sides of the nanowires, but their magnitudes differ. The difference manifesting in the spin textures when $x<-w/2$ and $x>w/2$ [Fig.~\ref{Spinhall}(c) and (d)] explains that it is related to the different excitation efficiency at the two sides of the magnets. This phenomenon is closely associated with the direction of the saturation magnetization: upon reversal, i.e., ${\bf M}_s\parallel -\hat{\bf y}$ as in Fig.~\ref{Spinhall}(b), the asymmetric distribution at the two sides of the magnet is switched. It is related to the unidirectionality of the electromagnetic field, noting the field propagates along the negative (positive) $\hat{\bf x}$-direction when ${\bf M}_s\parallel \hat{\bf y}$ (${\bf M}_s\parallel -\hat{\bf y}$). Turning off this unidirectionality by $\zeta^2=0$ via taking $d\ll w$ switches off the asymmetric distribution of the Hall response (refer to Appendix~\ref{appendix_chirality} for a detailed discussion). 

\section{Unidirectional longitudinal spin current}

\label{longitudinal_spin_pumping}

\begin{figure}[htbp]
    \centering
      \begin{minipage}[b]{0.492\columnwidth}
        \centering
        \includegraphics[width=\linewidth, clip, trim=0cm 0cm 0cm 0cm]{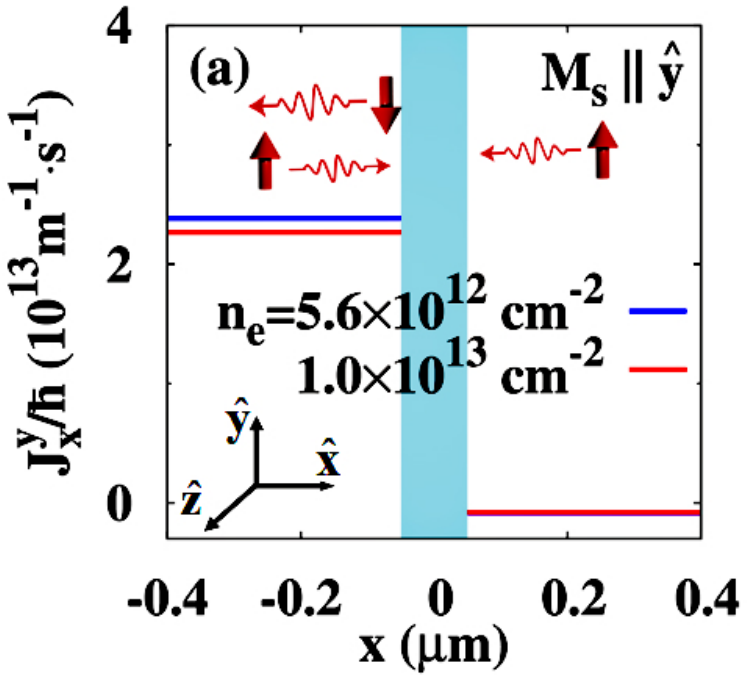}
    \end{minipage}
    \begin{minipage}[b]{0.492\columnwidth}
        \centering
        \includegraphics[width=\linewidth, clip, trim=0cm 0cm 0cm 0cm]{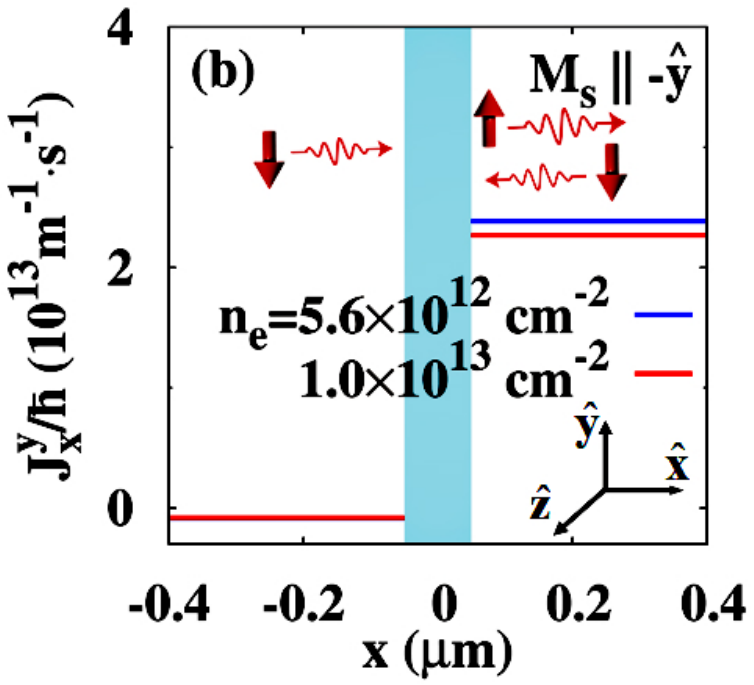}
    \end{minipage}
    \caption{Unidirectional longitudinal spin current excited by a near chiral electromagnetic field. (a) and (b) illustrate the magnitude and spatial distribution of the longitudinal spin current density for ${\bf M}_s\parallel\hat{\bf y}$ and ${\bf M}_s\parallel -\hat{\bf y}$ with different electron densities.}
    \label{current}
\end{figure}

The most pronounced effect of the unidirectionality in the near electromagnetic field is that it leads to a unidirectional flow of the longitudinal spin current along the field gradient $\hat{\bf x}$-direction. The unidirectionality of the magnetic field breaks the reciprocity in the photon emission $\xi=``+"$ and absorption $\xi=``-"$ processes. According to Eq.~\eqref{nonlinear_spin_current}, the pumped longitudinal spin current ${\pmb{\cal J}}_{x}^{L,y}(x)$ is polarized along $({\bf H}^{-\xi}\times{\bf H}^\xi)||\hat{\bf y}$. When ${\bf M}_s\parallel \hat{\bf y}$, a comparison between Eqs.~\eqref{nonlinear_spin_current} and \eqref{nonlinear_charge_current} leads to ${\pmb{\cal J}}_{x}^{L,y}(x>w/2)=(\hbar/2e){\bf J}^{e}_{x}(x>w/2)$ at the right-hand side of the magnet, indicating that the spin current is actually a charge current with spin polarization along $\hat{\bf y}$; while at the left-hand side of the magnet, the current flows oppositely with opposite spin polarization, which is not equal in magnitude that also gives rise to a net charge current.  The features of the unidirectional longitudinal spin current are summarized in Fig.~\ref{current}(a) and (b) with opposite directions of saturation magnetization. 

\section{Unidirectional charge current pumped by electromagnetic fields }
\label{charge_pumping}

\begin{center}
\begin{figure}[htp!]
\centering
\includegraphics[width=0.493\columnwidth,trim=0cm 1cm 0cm 0cm, clip]{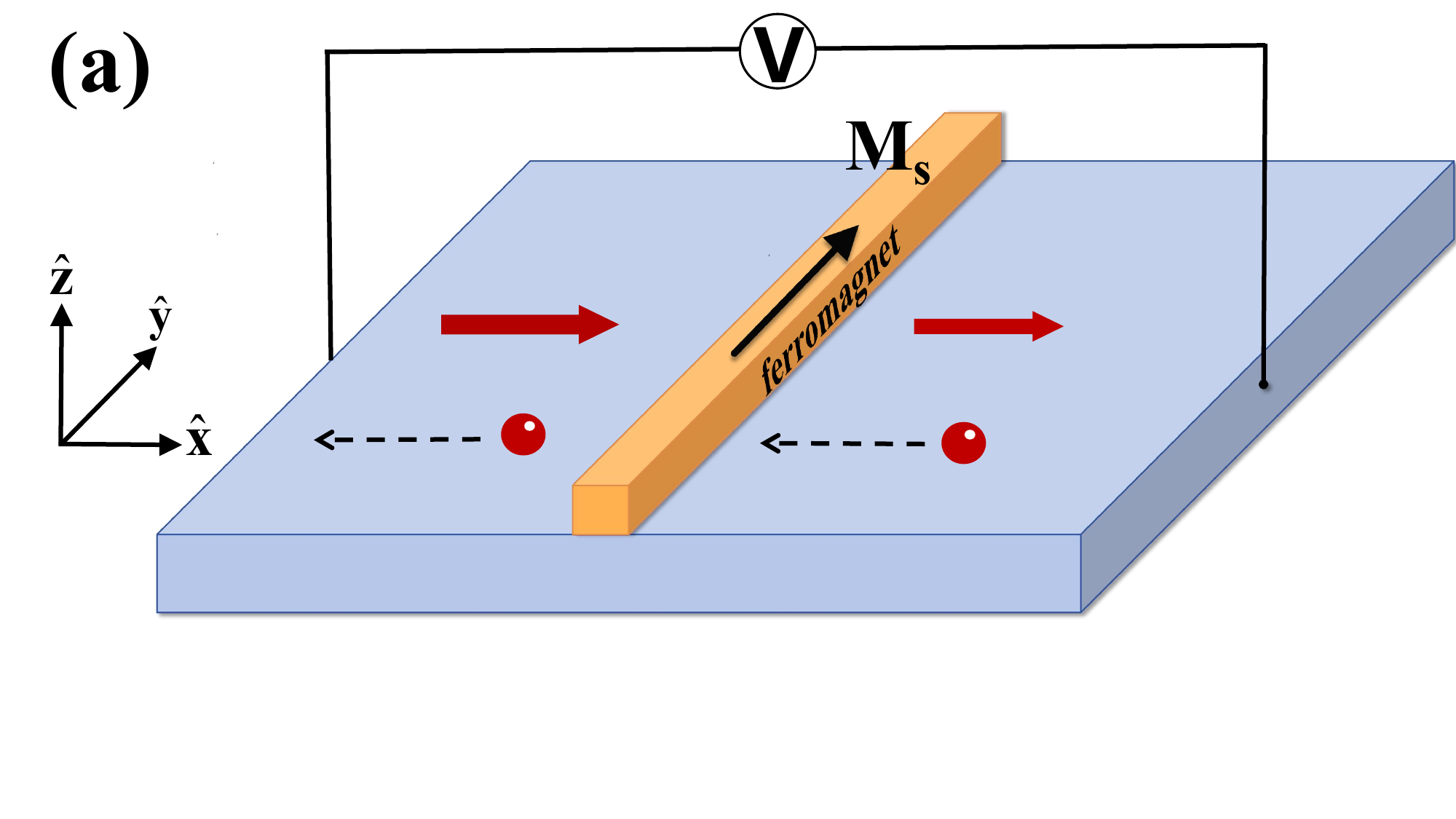}
\includegraphics[width=0.493\columnwidth,trim=0cm 1cm 0cm 0cm, clip]{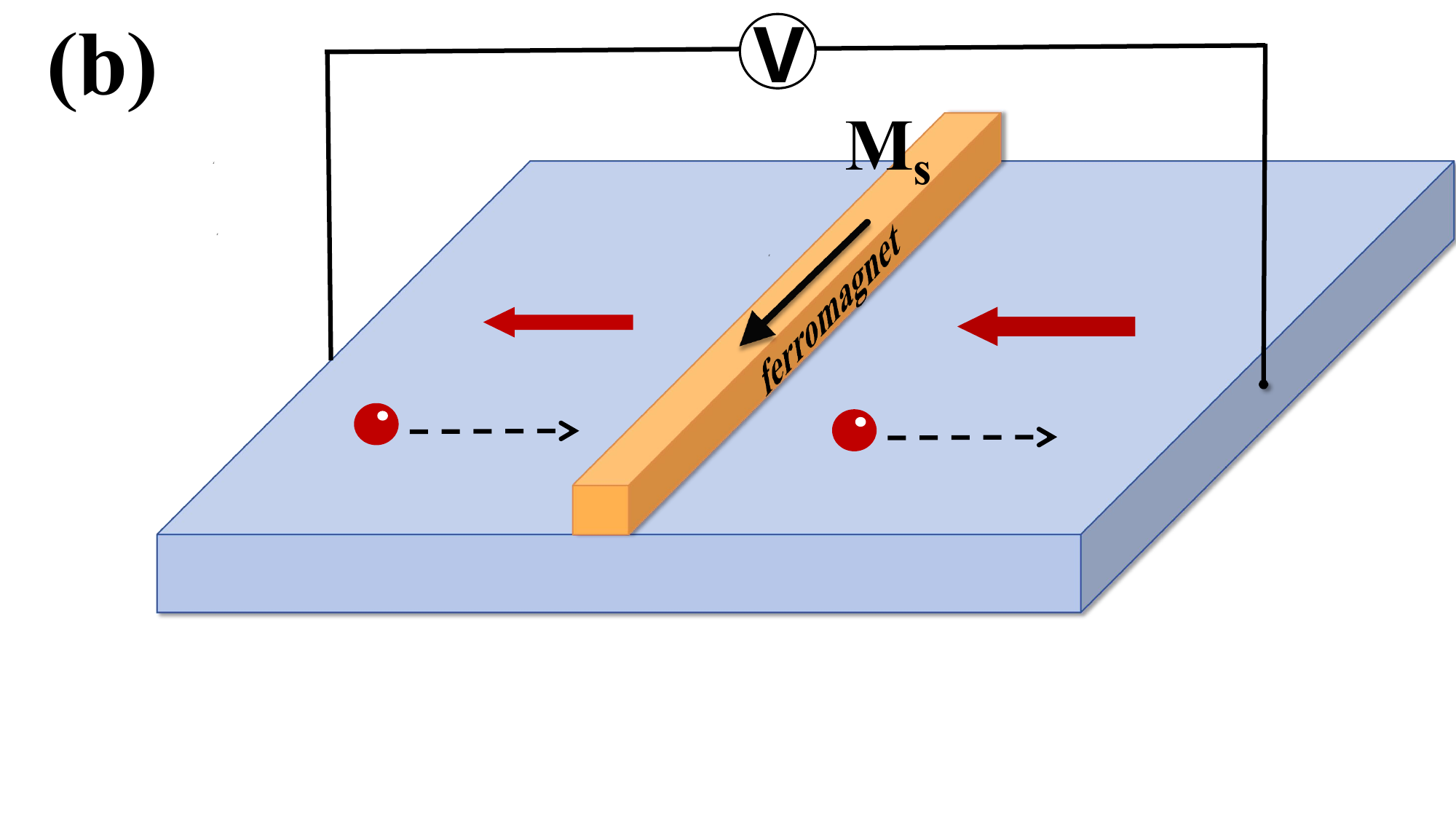}
\hspace{-1.3cm}\includegraphics[width=0.445\columnwidth,trim=0cm 0cm 0cm 0cm, clip]{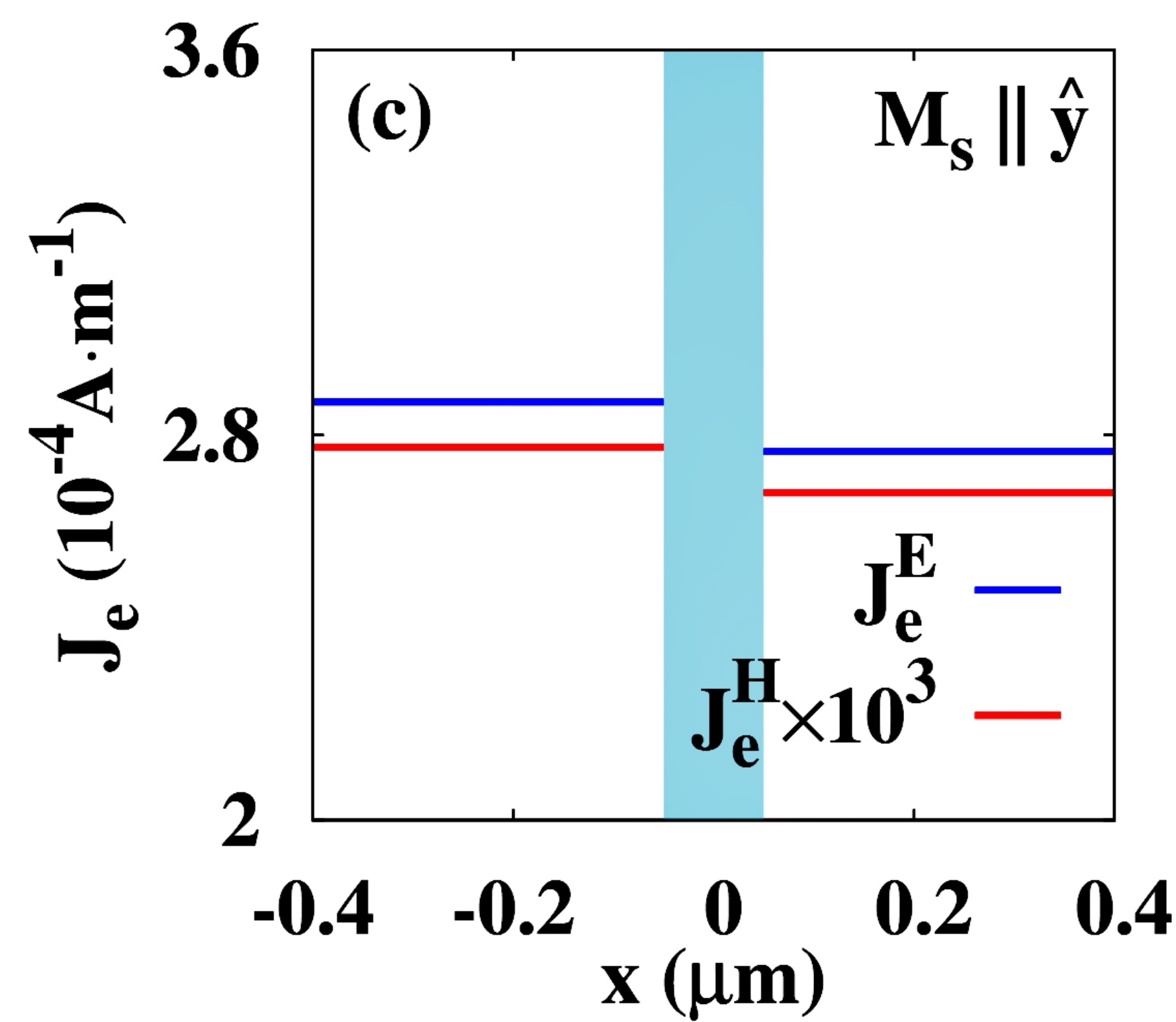}
\hspace{0.7cm}
\includegraphics[width=0.445\columnwidth,trim=0cm 0cm 0cm 0cm, clip]{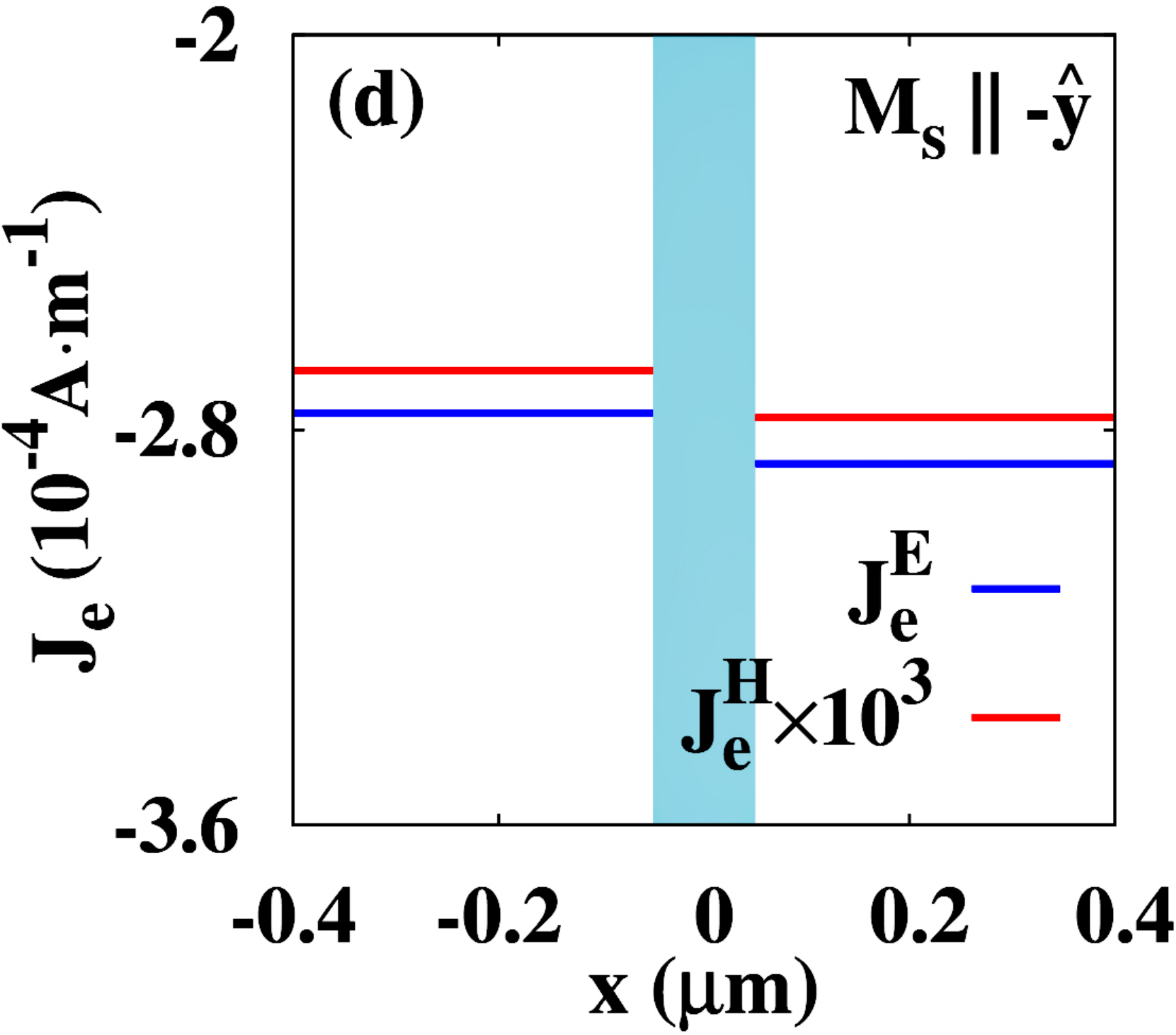}
\caption{(a) and (b) propose the measurement of the DC charge current pumped by the near electromagnetic field when the saturation magnetization ${\bf M}_s \parallel\ \hat{\bf y}$ and ${\bf M}_s\parallel-\hat{\bf y}$, respectively. The black dashed arrows denote the direction of electron flow (carrying charge $-\abs{e}$) and the red thick arrow denotes the charge current direction. (c) and (d) illustrate the spatial distribution of charge current pumped by the electric field  (the blue curve) and the magnetic field  (the red curve). The contribution from the electric field dominates. }
\label{ecurrent}
\end{figure}
\end{center}

An important by-product of the evanescent spin pumping is the generation of unidirectional charge current by the near magnetic field of the magnetic wires, which can be experimentally observed by a voltage, as illustrated in Fig.~\ref{ecurrent}(a) and (b) with ${\bf M}_s\parallel \hat{\bf y}$ and ${\bf M}_s\parallel -\hat{\bf y}$. Reversing the direction of the saturation magnetization switches the flow of the charge current or the sign of the voltage.

Here we compare the longitudinal unidirectional spin current and the charge current pumped by the stray magnetic field with $\zeta^2=1$.
The DC spin current  and charge current at the right-hand side of the magnetic wire read 
\begin{widetext}
\begin{subequations}
\begin{align}
    &{\pmb {\cal J}}^{y}_x\left(x>\frac{w}{2}\right) =\sum_{{\bf q}}\sum_{\xi=\pm}\frac{m\mu_0^2\gamma_e^2}{64\pi^2A}\frac{1}{q_\xi}\hat{\bf x}\otimes \hat{\bf y} [f(\varepsilon_q)-f(\varepsilon_q-\xi\hbar\omega)]\left(1-e^{-\abs{q_\xi-q_x}d}\right)^2\left(\frac{\sin ((q_\xi-q_x)w/2)}{q_\xi-q_x}\right)^2\nonumber\\
    &\times(1+\xi{\rm sgn}(q_\xi-q_x))^2M_z^2{\rm sgn}(q_\xi-q_x)+{\rm H.c.}\nonumber\\
    &=\sum_{q_x<-q_\omega}\sum_{q_y}\frac{m\mu_0^2\gamma_e^2}{8\pi^2A}\frac{1}{q_+} \hat{\bf x}\otimes \hat{\bf y}[f(\varepsilon_q)-f(\varepsilon_q-\hbar\omega)](1-e^{-\abs{q_+-q_x}d})^2\left(\frac{\sin ((q_+-q_x)w/2)}{q_+-q_x}\right)^2M_z^2
    \label{J^H_s},\\
    &{\bf J}^{e}_x\left(x>\frac{w}{2}\right)=\sum_{{\bf q}}\sum_{\xi=\pm}\frac{em\mu_0^2\gamma_e^2}{32\pi^2\hbar A}\frac{1}{q_\xi}\hat{\bf x}[f(\varepsilon_q)-f(\varepsilon_q-\xi\hbar\omega)](1-e^{-\abs{q_\xi-q_x}d})^2\left(\frac{\sin((q_\xi-q_x)w/2)}{q_\xi-q_x}\right)^2\nonumber\\
    &\times\left(1+\xi{\rm sgn}(q_\xi-q_x)\right)^2M_z^2+{\rm H.c.}\nonumber\\
   &=\sum_{q_x<-q_\omega}\sum_{q_y}\frac{em\mu_0^2\gamma_e^2}{4\pi^2\hbar A}\frac{1}{q_+}\hat{\bf x}[f(\varepsilon_q)-f(\varepsilon_q-\hbar\omega)](1-e^{-\abs{q_+-q_x}d})^2\left(\frac{\sin((q_+-q_x)w/2)}{q_+-q_x}\right)^2M_z^2. 
   \label{J^H_e}
\end{align}
\end{subequations}
\end{widetext}
Both arise solely from the photon emission $\xi =``+"$ scattering process, which contributes to the negative spin current and positive charge current.  The relation ${\pmb {\cal J}}^{y}_x(x>w/2)=(\hbar/2e){\bf J}^{e}_x(x>w/2)$  indicates that the spin current is actually a charge current with spin polarization along the photon spin $\hat{\bf y}$-direction.

At the left-hand side of the magnetic wire, both the photon emission and absorption processes with $\xi=``\pm"$ contribute to the spin current and charge current.
For the photon emission $\xi=``+"$ process,
\begin{widetext}
\begin{subequations}
\begin{align}
   {\pmb {\cal J}}^{y}_{x,+}\left(x<-\frac{w}{2}\right)&=-\frac{m\mu_0^2\gamma_e^2}{8\pi^2A}\sum_{q_x<-q_\omega}\sum_{q_y}\frac{\hat{\bf x}\otimes \hat{\bf y}}{q_+} [f(\varepsilon_q)-f(\varepsilon_q-\hbar\omega)](1-e^{-\abs{q_++q_x}d})^2\left(\frac{\sin (q_++q_x)w/2}{q_++q_x}\right)^2M_z^2,\\
{\bf J}^{e}_{x,+}\left(x<-\frac{w}{2}\right)&=-\frac{em\mu_0^2\gamma_e^2}{4\pi^2\hbar A}\sum_{q_x<-q_\omega}\sum_{q_y}\frac{\hat{\bf x}}{q_+}[f(\varepsilon_q)-f(\varepsilon_q-\hbar\omega)](1-e^{-\abs{q_++q_x}d})^2\left(\frac{\sin((q_++q_x)w/2)}{q_++q_x}\right)^2M_z^2.
    \label{current+}
\end{align}
\end{subequations}
For the photon absorption $\xi=``-"$ process,
\begin{subequations}
\begin{align}
{\pmb {\cal J}}^{y}_{x,-}\left(x<-\frac{w}{2}\right)&=\frac{m\mu_0^2\gamma_e^2}{8\pi^2A}\sum_{q_x,q_y}\frac{\hat{\bf x}\otimes \hat{\bf y}}{q_-}[f(\varepsilon_q)-f(\varepsilon_q+\hbar\omega)](1-e^{-\abs{q_-+q_x}d})^2\left(\frac{\sin (q_-+q_x)w/2}{q_-+q_x}\right)^2M_z^2,\\
{\bf J}^{e}_{x,-}\left(x<-\frac{w}{2}\right)&=-\frac{em\mu_0^2\gamma_e^2}{4\pi^2\hbar A}\sum_{q_x,q_y}\frac{\hat{\bf x}}{q_-}[f(\varepsilon_q)-f(\varepsilon_q+\hbar\omega)](1-e^{-\abs{q_-+q_x}d})^2\left(\frac{\sin((q_-+q_x)w/2)}{q_-+q_x}\right)^2M_z^2.
\label{current-}
\end{align}
\end{subequations}
\end{widetext}
It is seen that  $ {\pmb {\cal J}}^{y}_{x,+}=({\hbar}/{2e}) {\bf J}^{e}_{x,+} $ and $ {\pmb {\cal J}}^{y}_{x,-}=-({\hbar}/{2e}) {\bf J}^{e}_{x,-}$. 
So in the region $x<-w/2$, both photon emission and absorption processes with $\xi=``\pm"$  contribute constructively to a positive spin current  ${\pmb {\cal J}}^{y}_x$, which flows in the direction opposite to the charge current ${\bf J}^{e}_x$. This reflects the underlying spin-momentum locking, wherein electrons propagating in the opposite direction $(\pm\hat{\bf x})$ carry opposite-oriented spins $(\mp\hat{\bf y})$.

Notably, according to Eqs.~\eqref{current+} and \eqref{current-}, the counterpropagating charge  currents  originating from the photon emission and  absorption $\xi=``\pm"$
 scattering processes do not fully cancel, giving rise to a finite net charge current. The pumped current is dominated by the electric field, as illustrated in Fig.\eqref{ecurrent}(c) and (d).
At the right-hand side of the magnetic wire, the charge current pumped by the electric field
\begin{align}
   &{\bf J}^{e,\rm E}_x\left(x>\frac{w}{2}\right)
    =\frac{e^3\mu_0^2\hat{\bf x}}{2\pi^2m\hbar A}\sum_{q_x<-q_\omega}\sum_{q_y}\frac{q_y^2}{q_+}\left[f(\varepsilon_q)-f(\varepsilon_q-\hbar\omega)\right]\nonumber\\
    &\times
    (1-e^{-\abs{q_+-q_x}d})^2\left(\frac{\sin ((q_+-q_x)w/2)}{(q_+-q_x)^2}\right)^2M_z^2
\end{align}
is solely contributed by the photon emission  $\xi=``+"$ process. 
While at the left-hand side of the magnetic wire, the charge current  
\begin{align}
     &{\bf J}^{e,\rm E}_x\left(x<-\frac{w}{2}\right)=-\frac{e^3\mu_0^2}{16\pi^2m\hbar A}\sum_{{\bf q}}\sum_{\xi=\pm}\frac{q_y^2}{q_\xi}\hat{\bf x}\nonumber\\
    &\times[f(\varepsilon_q)-f(\varepsilon_q-\xi\hbar\omega)](1-e^{-\abs{q_\xi+q_x}d})^2\nonumber\\
    &\times\left(\frac{\sin (q_\xi+q_x)w/2}{(q_\xi+q_x)^2}\right)^2(1+\xi{\rm sgn}(-q_\xi-q_x))^2M_z^2+{\rm H.c.}
\end{align}
is contributed by both the photon emission $\xi=``+"$ and absorption $\xi=``-"$ processes. This explains the asymmetric distribution of the charge current at the two sides of the magnetic wire. Upon reversing ${\bf M}_s$, i.e., ${\bf M}_s \parallel -\hat{\bf y}$ as in Fig.~\eqref{ecurrent}(d), the DC charge current switches its flow direction.

\section{Discussion and conclusion}
\label{conclusion}

In conclusion, we have demonstrated a transverse spin pumping that is distinct from the conventional longitudinal one, which does not originate from the transfer of the magnon/photon spin but a joint effect of the electric and magnetic fields. It holds a high efficiency: The pumped Hall spin current reaches a magnitude corresponding to a spin conductivity $\sigma_s=0.3\abs{e}/(4\pi)$ under an electric field $1~{\rm kV/cm}$, even larger than the predicted value $0.2\abs{e}/(4\pi)$ due to the spin Hall effect in  $\rm MoS_2$~\cite{conductivity_Mos2_1,JS,conductivity_Mos2_2,conductivity_Mos2_3,conductivity_Mos2_4}. The chirality of the electromagnetic field is reflected in the spatial distribution of the spin current, which gives rise to unidirectional transverse spin and longitudinal spin/charge currents due to nonreciprocal photon emission and absorption, which can be switched by the magnetization direction. Our predictions can be tested in future spintronic devices by exploiting nanomagnets and optical fields.

\begin{acknowledgments}
This work is financially supported by the National Key Research and Development Program of China under Grant No.~2023YFA1406600 and the National Natural Science Foundation of China under Grant No.~12374109. 
\end{acknowledgments}

\begin{appendix}

\section{Conditions of transverse spin pumping by one-dimensional electromagnetic fields}

\label{appendix_conditions}

We demonstrate here that the emergence of Hall-type/transverse spin pumping by one-dimensional optical fields relies on a nontrivial phase difference between the magnetic and electric components of the driving fields, distinct from the two-dimensional scenario.

To this end, we consider a one-dimensional optical field with Fourier components propagating spatially along the $\hat{\bf x}$-direction, i.e.,  ${\bf E}({\bf k})=2\pi\delta(k_y){\bf E}(k_x)$ and ${\bf H}({\bf k})=2\pi\delta(k_y){\bf H}(k_x)$. 
Here, the directions of the electric and magnetic components are set to be relatively arbitrary, such that the DC transverse spin current reads
\begin{widetext}
\begin{align}
\pmb{\cal J}_s^{\nu}({\pmb\rho})
&=\frac{ie\mu_0\gamma_e\hbar^4}{8m^2\omega A^2}\sum_{{\bf q}}\sum_{\xi=\pm}\sum_{k_x^\prime,k_x}\xi e^{i(k^\prime_x-k_x)x}\frac{f(\varepsilon_{k_x,q_y})-f(\varepsilon_{q})}{\left(\varepsilon_{q}-\varepsilon_{k_x^\prime,q_y}-\xi\hbar\omega+i\delta\right)\left(\varepsilon_{k_x,q_y}-\varepsilon_{ q}+\xi\hbar\omega+i\delta\right)}\nonumber\\
&\times \bigg{[}\left(k^\prime_x(q_x+k_x)E_x^\xi(q_x-k_x)\hat{\bf x}+2q^2_y E_y^{\xi}(q_x-k_x)\hat{\bf y}\right)\otimes{H}_\nu^{-\xi}(k^\prime_x-q_x)\hat{\pmb \nu}\nonumber\\
&-\left(k^\prime_x(k^\prime_x+q_x){E}_x^{-\xi}(k^\prime_x-q_x)\hat{\bf x}+2q_y^2E_y^{-\xi}(k_x^\prime-q_x)\hat{\bf y}\right)\otimes{H}_\nu^{\xi}(q_x-k_x)\hat{\pmb\nu}\bigg{]}+{\rm H.c.},
\label{spin_hall_one_dimensional}
\end{align}
in which $\hat{\pmb \nu}$ is the spin-polarization direction determined solely by the direction of the  AC magnetic field, and $\{\hat{\bf x},{\hat{\bf y}}\}$ are the electron flow directions relying on the directions of the electric field.
With the contour integral, the Hall-type spin current at $x>0$ is reduced to
\begin{align}
{\pmb{\cal J}}^{\nu}_s(x>0)&=\frac{e\mu_0\gamma_e}{4\omega A}\sum_{{\bf q}}\sum_{\xi=\pm} i\xi [f(\varepsilon_q)-f(\varepsilon_q-\xi\hbar\omega)]\nonumber\\
&\times\frac{1}{q_\xi^2}\left[q_\xi(q_x+q_\xi)E_x^\xi(q_x-q_\xi)\hat{\bf x}+2q_y^2E_y^\xi(q_x-q_\xi)\hat{\bf y}\right]\otimes H_\nu^{-\xi}(q_\xi-q_x)\hat{\pmb\nu}+{\rm H.c.},
\label{General Hall spin current}
\end{align}
where $q_\xi=\sqrt{q^2_x-2\xi m\omega/\hbar}$. ${\pmb{\cal J}}^{\nu}_s(x<0)$ is obtained by taking $q_\xi\rightarrow-q_\xi$.

It shows that a finite spin current $\pmb{\cal J}^\nu_s$ can be generated when $E^{\xi}_xH^{-\xi}_\nu$ or $E^{\xi}_yH^{-\xi}_\nu$ is \textit{not purely real}, indicating that the Hall spin current can only be driven when the electric and magnetic fields are neither in phase nor out of phase under the one-dimensional driving.  Two representative examples are addressed as follows.

\subsubsection{Linearly polarized electromagnetic field}

We consider a focused linearly polarized electromagnetic field with Fourier components propagating spatially along $\hat{\bf x}$, in which the electric and magnetic fields are polarized along the unit vectors $\hat{\bf n}_{\rm E}$ and $\hat{\bf n}_{\rm H}$, respectively.  
The fields are then given by
\begin{align}
    {\bf E}({\bf r},t)&=\sum_{k_x}E_0(k_x,z=0) e^{ik_xx-i\omega t}\hat{\bf n}_{\rm E}+{\rm H.c.},\nonumber\\
    {\bf H}({\bf r},t)&=\sum_{k_x}H_0(k_x,z=0)e^{i\phi}e^{ik_xx-i\omega t}\hat{\bf n}_{\rm H}+{\rm H.c.},
\end{align}
where $E_0(k_x)$ and $H_0(k_x)$ are the amplitudes of the electric and magnetic fields and $\phi$ denotes the phase difference between them.   According to Eq.~\eqref{form_expression}, the Fourier components of electric and magnetic field are written as  $ {\bf E}^\xi({\bf k})=2\pi\delta(k_y) E_0(\xi k_x)\hat{\bf n}_{\rm E}$ and ${\bf H}^\xi({\bf k})=2\pi\delta(k_y) H_0(\xi k_x)e^{i\xi\phi}\hat{\bf n}_{\rm H}$.
Substitution of these Fourier components
into Eq.~\eqref{General Hall spin current} yields  the transverse spin current 
\begin{align}
    {\pmb{\cal J}}^{\nu}_s(x>0)&=\frac{e\mu_0\gamma_e}{4\omega A}\sum_{{\bf q}}\sum_{\xi=\pm}\xi\frac{1}{q_\xi^2}E_0(\xi(q_x-q_\xi))H_0(\xi(q_x-q_\xi)) [f(\varepsilon_q)-f(\varepsilon_q-\xi\hbar\omega)]\nonumber\\
&\times\left[q_\xi(q_x+q_\xi){n}_{{\rm E},x}\hat{\bf x}+2q_y^2{{ n}}_{{\rm E}, y}\hat{\bf y}\right]\otimes {n}_{{\rm H},\nu}\hat{\pmb \nu}\left(ie^{-i\xi\phi}+{\rm H.c.}\right),
\end{align}
where $ n_{{\rm E},x}\hat{\bf x}$ $(n_{{\rm E},y}\hat{\bf y})$ is the $x\mbox{-}$ $(y\mbox{-})$ component of $\hat{\bf n}_{\rm E}$ and $ n_{{\rm H},\nu}\hat{\pmb\nu}$ is the $\nu$-component of $\hat{\bf n}_{\rm H}$, respectively.
From the factor $\left(ie^{-i\xi\phi}+{\rm H.c.}\right)$, we find that only the phase difference $\phi\neq 0$ or $\pi$, the Hall spin current exists.

\subsubsection{Circularly polarized electromagnetic field}

Further, we consider a circularly polarized  optical beam with a Gaussian distribution along the $\hat{\bf x}$-direction and a translational symmetry along the $\hat{\bf y}$-direction, adapted from Eq.~\eqref{gaussian beam}, 
\begin{align}
   {\bf  E}({\pmb\rho},z=0)&=E_0e^{-{x^2}/{w^2_0}}e^{-i\omega t}\left(\begin{array}{ccc}
         i  \\
         1\\
         0
    \end{array}\right)+{\rm H.c.},\nonumber\\
   {\bf H}({\pmb\rho},z=0)&=\frac{ E_0}{\mu_0c}e^{-{x^2}/{w^2_0}}e^{-i\omega t}\left(\begin{array}{ccc}
         1  \\
         -i\\
         0
    \end{array}\right)+{\rm H.c.}\label{polarized-eh}.
\end{align}
According to Eq.~\eqref{form_expression}, the Fourier components $ {\bf  E}^\xi({\bf k})=2\pi\delta(k_y)E(\xi k_x)e^{-i\xi\omega t}(i\xi,1,0)^T $ and $ {\bf H}^\xi({\bf k})=2\pi \delta(k_y)   H(\xi k_x)(1,-i\xi,0)^T$, in which $E(\xi k_x)= E_0\sqrt{\pi w^2_0}e^{-{\xi k_xw^2_0}/{4}}$ and $H(\xi k_x)= [E_0/(\mu_0c)]\sqrt{\pi w^2_0}e^{-{\xi k_xw^2_0}/{4}}$.
So the Hall spin current becomes 
\begin{align}
    {\pmb{\cal J}}_s(x>0)&=\frac{e\mu_0\gamma_e}{4\omega A}\sum_{{\bf q}}\sum_{\xi=\pm}\frac{1}{q_\xi^2}E(\xi(q_x-q_\xi))H(\xi(q_x-q_\xi)) \left[f(\varepsilon_q)-f(\varepsilon_q-\xi\hbar\omega)\right]\nonumber\\
&\times\left[q_\xi(q_x+q_\xi)i\xi\hat{\bf x}+2q_y^2\hat{{\bf y}}\right]\otimes \left(i\xi\hat{\bf x}-\hat{\bf y}\right)+{\rm H.c.}.
\end{align}
A $\pi/2$ phase difference exists between the electric and the magnetic field components in both $x$ and $y$ directions, which leads to the Hall spin current 
\begin{align}
    {\pmb{\cal J}}^x_x&=-\sum_{\bf q}\sum_{\xi=\pm}\frac{e\mu_0\gamma_e\pi w_0^2}{4\omega\mu_0cA}\frac{q_x+q_\xi}{q_\xi}e^{-\xi(q_x-q_\xi)w^2_0/2}[f(\varepsilon_q)-f(\varepsilon_q-\xi\hbar\omega)]\hat{\bf x}\otimes\hat{\bf x}+{\rm H.c.},\nonumber\\
    {\pmb{\cal J}}^y_y&=-\sum_{\bf q}\sum_{\xi=\pm}\frac{e\mu_0\gamma_e\pi w_0^2}{2\omega\mu_0cA}\frac{q^2_y}{q^2_\xi}e^{-\xi(q_x-q_\xi)w^2_0/2}[f(\varepsilon_q)-f(\varepsilon_q-\xi\hbar\omega)]\hat{\bf y}\otimes\hat{\bf y}+{\rm H.c.}.
\end{align}
Therefore, both components in the Hall spin current can be driven by an electromagnetic field with a $\pi/2$ phase difference between the associated components of the electric and magnetic fields.

\section{Transverse spin pumping by a focused laser}
\label{appendix_optical_pumping}

As addressed in the main text, a local and time-varying electromagnetic field ${\bf E}({\pmb \rho},t)=\sum_{\bf q}({\bf E}^+({\bf q})e^{-i\omega t}+{\bf E}^-({\bf q})e^{i\omega t})e^{i{\bf q}\cdot {\pmb \rho}}$ and  ${\bf H}({\pmb \rho},t)=\sum_{\bf q}({\bf H}^+({\bf q})e^{-i\omega t}+{\bf H}^-({\bf q})e^{i\omega t})e^{i{\bf q}\cdot {\pmb \rho}}$ of frequency $\omega$ can pump a DC spin current into the two-dimensional electron gas (2DEG) according to 
\begin{align}
    {\pmb {\cal J}}^{\rm DC}_{s}(\boldsymbol{\rho})  &=\frac{\hbar^2}{4mA}\sum_{{\bf k},{\bf k}^\prime,{\bf q}}\sum_{\xi=\pm}\sum_{\alpha,\beta,\gamma=\{\uparrow\downarrow\}}
\frac{ (f(\varepsilon_{q})-f(\varepsilon_{k}))e^{i({\bf k}^\prime-{\bf k})\cdot\boldsymbol{\rho}}}{\left(\varepsilon_{ k'}-\varepsilon_{ q}+\xi\hbar\omega-i\delta\right)\left(\varepsilon_{ k}-\varepsilon_{ q}+\xi\hbar\omega+i\delta\right)}\left({\pmb\sigma}_{\beta\alpha}\otimes{\bf k}^\prime\right)\nonumber\\
&\times\mathcal{G}^{-\xi}_{\alpha\gamma}({\bf k}^\prime,{\bf q})\mathcal{G}^\xi_{\gamma\beta}({\bf q},{\bf k})+{\rm H.c.},
\label{nonlinear_approximation_current}
\end{align}
in which the matrix $
\mathcal{G}^\xi({\bf k},{\bf k}^\prime)=[i\xi e\hbar/(2m\omega A)]({\bf k}+{\bf k}^\prime)\cdot{\bf E}^\xi({\bf k}-{\bf k}^\prime)+[{\mu_0\gamma_e\hbar}/({2A})]{\bf H}^\xi({\bf k}-{\bf k}^\prime)\cdot{\pmb\sigma}$ in the spin space describes the interaction between electrons of different wave vectors ${\bf k}$ and ${\bf k}'$,  mediated via the electromagnetic radiation.
Here, $A$ denotes the area of the crystal, $\boldsymbol{\sigma}$ is the Pauli matrix, $e=-|e|$ represents the electron charge, $m$ is the effective electron mass, $\mu_0$ is the vacuum permeability, and $\gamma_e$ denotes the effective gyromagnetic ratio of electrons. 
The DC spin current may be contributed by the terms $|{\bf E}|^2$, $|{\bf E}||{\bf H}|$, and $|{\bf H}|^2$, which contain both the longitudinal and transverse components. Here, the longitudinal (transverse) spin current is along (normal to) the gradient of the field.
After summing over the spin indices, the $|{\bf E}|^2$-term has no contribution, such that the DC spin current becomes
\begin{align}
{\pmb {\cal J}}^{\rm DC}_{s}(\boldsymbol{\rho}) 
&=\frac{i\hbar^4\mu^2_0\gamma^2_e}{8mA^3}\sum_{{\bf k},{\bf k}^\prime,{\bf q}}\sum_{\xi=\pm}
\frac{ \left(f(\varepsilon_{q})-f(\varepsilon_{k})\right)e^{i({\bf k}^\prime-{\bf k})\cdot\boldsymbol{\rho}}}{\left(\varepsilon_{ k'}-\varepsilon_{ q}+\xi\hbar\omega-i\delta\right)\left(\varepsilon_{ k}-\varepsilon_{ q}+\xi\hbar\omega+i\delta\right)}\left({\bf H}^{-\xi}({\bf k}'-{\bf q})\times{\bf H}^{\xi}({\bf q}-{\bf k})\right)\otimes{\bf k}^\prime \nonumber\\
&-\frac{i\hbar^4e\mu_0\gamma_e}{8m^2\omega A^3}\sum_{{\bf k},{\bf k}^\prime,{\bf q}}\sum_{\xi=\pm}
\frac{ \xi\left(f(\varepsilon_{q})-f(\varepsilon_{k})\right)e^{i({\bf k}^\prime-{\bf k})\cdot\boldsymbol{\rho}}}{\left(\varepsilon_{ k'}-\varepsilon_{ q}+\xi\hbar\omega-i\delta\right)\left(\varepsilon_{ k}-\varepsilon_{ q}+\xi\hbar\omega+i\delta\right)}\left(({\bf k}'+{\bf q})\cdot{\bf E}^{-\xi}({\bf k}'-{\bf q})\right)\nonumber\\
&\times{\bf H}^{\xi}({\bf q}-{\bf k}) \otimes({\bf k}+{\bf k}')+{\rm H.c.},
\end{align}
in which the first term, proportional to the spin of photons ${\bf H}\times{\bf H}^*$, originates from the transfer of photonic spin into the electrons, while the second term describes the spin current driven jointly by the electric and magnetic fields. Here, we mainly focus on the latter contribution due to its high efficiency.

When the driven electromagnetic field comes from a focused laser beam, it is convenient to adopt the polar coordinate system in the calculation, i.e., the wave vector ${\bf k}=k(\cos\theta_{\bf k}\hat{\bf x}+\sin\theta_{\bf k}\hat{\bf y})$, ${\bf k}'\rightarrow\{k',\theta_{\bf k'}\}$, ${\bf q}\rightarrow\{q,\theta_{\bf q}\}$, and ${\pmb \rho}\rightarrow\{\rho,\theta_{\boldsymbol{\rho}}\}$.
Performing the contour integral surrounding the residues $k=q_\xi-i\delta$ and $k'=q_\xi+i\delta$ in the complex plane, in which $q_\xi$ obeys the energy conservation according to $\varepsilon_{q_\xi}=\varepsilon_q-\xi\hbar\omega$, the driven current by the joint effect of electric and magnetic fields reads
\begin{align}
{\pmb {\cal J}}^{\rm EH}_{s}(\boldsymbol{\rho})&\approx-\sum_{\xi=\pm}\frac{i\xi e\mu_0\gamma_e}{8\omega (2\pi)^4}\int_0^\infty qdq\int_0^{2\pi}d\theta_{\bf q}\int_{\theta_{\pmb{\rho}}-\pi/2}^{\theta_{\pmb{\rho}}+\pi/2}d\theta_{\bf k}d{\theta_{\bf k'}}
\left(f(\varepsilon_{q})-f(\varepsilon_{q}-\xi\hbar\omega)\right)\nonumber\\&
\times \left(({\bf q}^\prime_\xi+{\bf q})\cdot{\bf E}^{-\xi}({\bf q}^\prime_\xi-{\bf q})\right){\bf H}^{\xi}({\bf q}-{\bf q}_\xi) \otimes({\bf q}^\prime_\xi+{\bf q}_\xi)e^{i({\bf q}^\prime_\xi-{\bf q}_\xi)\cdot\boldsymbol{\rho}}+{\rm H.c.},
\label{J_EH}
\end{align}
in which ${\bf q}_\xi\rightarrow\{q_\xi,\theta_{\bf k}\}$ and ${\bf q}^\prime_\xi\rightarrow\{q_\xi,\theta_{\bf k'}\}$ are the wave vectors after performing the contour integral over ${\bf k}$ and ${\bf k'}$.

For a circularly polarized Gaussian beam that is irradiated normally on the 2DEG, the electromagnetic field interacting with electrons reads
\begin{align}
    &{\bf E}({\boldsymbol{\rho}},z,t)=E_0\frac{w_0}{w(z)}e^{-\frac{\rho^2}{w^2(z)}}e^{-i(kz+\omega t)}e^{i(\phi(z)-\frac{k\rho^2}{2R(z)})}\left(\begin{array}{c}
          1 \\i\\0
    \end{array}\right)+{\rm H.c.},\nonumber\\
    &{\bf H}({\boldsymbol{\rho}},z,t)=\frac{E_0}{\mu_0c}\frac{w_0}{w(z)}e^{-\frac{\rho^2}{w^2(z)}}e^{-i(kz+\omega t)}e^{i\left(\phi(z)-\frac{k\rho^2}{2R(z)}\right)}\left(\begin{array}{c}
          i \\-1\\0
    \end{array}\right)+{\rm H.c.},
    \label{gaussian beam}
\end{align}
where $E_0$ is the field amplitude, $w(z)$, $R(z)$, and $\phi(z)$ represent the beam radius with waist $w_0$ at $z=0$, wavefront radius of curvature, and phase of the Gaussian beam, respectively, all of which vary slowly with respect to the surface normal $z$ compared to the fast oscillation $e^{ikz}$. 
When the beam radius $w(z)\rightarrow0$ is sufficiently narrow  and neglecting the small radial phase variation $e^{-ik\rho^2/(2R(z))}$, the Gaussian beam approaches an ideal point source, i.e.,
\begin{align}
    &{\bf E}({\pmb{\rho}},z=0,t)\rightarrow E_0\delta(\rho)e^{-i\omega t}\left(\begin{array}{c}
          1 \\i\\0
    \end{array}\right)+{\rm H.c.},\nonumber\\
    &{\bf H}({\pmb{\rho}},z=0,t)\rightarrow \frac{E_0}{\mu_0c}\delta(\rho)e^{-i\omega t}\left(\begin{array}{c}
          i \\-1\\0
    \end{array}\right)+{\rm H.c.}.
\end{align}
Such electromagnetic field can be written as 
\begin{align}
    {\bf E}({\pmb\rho},t)
    &=\sum_{{\bf k}}\sum_{\xi=\pm}{\bf E}^\xi({\bf k})e^{-i\xi\omega t}e^{i{\bf k}\cdot\pmb\rho},\nonumber\\
    {\bf H}(\pmb\rho,t)&=\sum_{{\bf k}}\sum_{\xi=\pm}{\bf H}^\xi({\bf k})e^{-i\xi\omega t}e^{i{\bf k}\cdot \pmb\rho},
    \label{form_expression}
\end{align} in which the Fourier components ${\bf E}^\xi({\bf k})=E_0(1,i\xi,0)^T$  and ${\bf H}^{\xi}({\bf k})=[E_0/(\mu_0c)](i\xi,-1,0)^T$.

Substituting these Fourier components  into Eq.~\eqref{J_EH} yields the driven spin current
\begin{align}
&{\pmb {\cal J}}^{\rm EH}_{s}(\boldsymbol{\rho})=-\sum_{\xi=\pm}\frac{i\xi e\gamma_eE_0^2}{8\omega (2\pi)^4c}\int_0^\infty qdq\int_0^{2\pi}d\theta_{\bf q}\int_{\theta_{\pmb{\rho}}-\pi/2}^{\theta_{\pmb{\rho}}+\pi/2}d\theta_{\bf k}d{\theta_{\bf k'}}
\left(f(\varepsilon_{q})-f(\varepsilon_{q}-\xi\hbar\omega)\right)q_\xi^2e^{i({\bf q}^\prime_\xi-{\bf q}_\xi)\cdot\boldsymbol{\rho}}\nonumber\\&\times \left(\cos\theta_{\bf q}+\cos\theta_{\bf k'}-i\xi\sin\theta_{\bf q}-i\xi\sin\theta_{\bf k'}\right)(i\xi\hat{\bf x}-\hat{\bf y})\otimes\left[\hat{\bf x}(\cos\theta_{\bf k}+\cos\theta_{\bf k'})+\hat{\bf y}(\sin\theta_{\bf k}+\sin\theta_{\bf k'})\right]+{\rm H.c.}.
\end{align}
In the degenerate regime with the temperature $k_BT\ll \mu$ and with the field frequency $\hbar\omega\ll\mu$, 
\begin{align}
    {\pmb {\cal J}}^{\rm EH}_{s}(\boldsymbol{\rho})&\approx-\sum_{\xi=\pm}\frac{ime\gamma_eE_0^2q_F^2}{8\hbar\omega (2\pi)^4c}\int_0^{2\pi}d\theta_{\bf q}\int_{\theta_{\boldsymbol{\rho}}-\pi/2}^{\theta_{\boldsymbol{\rho}}+\pi/2}d\theta_{\bf k}d{\theta_{\bf k'}}(\cos\theta_{\bf k}+\cos\theta_{\bf k'}-i\xi\sin\theta_{\bf k}-i\xi\sin\theta_{\bf k'})(i\xi\hat{\bf x}-\hat{\bf y})\nonumber\\
    &\otimes\left[\hat{\bf x}(\cos\theta_{\bf k}+\cos\theta_{\bf k'})+\hat{\bf y}(\sin\theta_{\bf k}+\sin\theta_{\bf k'})\right]e^{iq_F\rho\cos(\theta_{\bf k'}-\theta_{\boldsymbol{\rho}})}e^{-iq_F\rho\cos(\theta_{\bf k}-\theta_{\boldsymbol{\rho}})}+{\rm H.c.}\nonumber\\
    &=-\frac{ime\gamma_eE_0^2q_F^2}{8\hbar (2\pi)^3c}\int_{\theta_{\boldsymbol{\rho}}-\pi/2}^{\theta_{\boldsymbol{\rho}}+\pi/2}d\theta_{\bf k}d{\theta_{\bf k'}}(\sin\theta_{\bf k'}\hat{\bf x}-\cos\theta_{\bf k'}\hat{\bf y})\otimes\left[\hat{\bf x}(\cos\theta_{\bf k}+\cos\theta_{\bf k'})+\hat{\bf y}(\sin\theta_{\bf k}+\sin\theta_{\bf k'})\right]\nonumber\\
    &\times e^{iq_F\rho\cos(\theta_{\bf k'}-\theta_{\boldsymbol{\rho}})}e^{-iq_F\rho\cos(\theta_{\bf k}-\theta_{\boldsymbol{\rho}})}+{\rm H.c.},
\end{align}
in which the Fermi wave vector $q_F$ satisfies $\hbar^2q_F^2/(2m)=\mu$.
In the radial coordinate with $\hat{\bf e}_{\rho}=\cos\theta_{\boldsymbol{{\rho}}}\hat{\bf x}+\sin\theta_{\boldsymbol{\rho}}\hat{\bf y}$ and $\hat{\bf e}_{\perp}=-\sin\theta_{\boldsymbol{{\rho}}}\hat{\bf x}+\cos\theta_{\boldsymbol{\rho}}\hat{\bf y}$,
\begin{align}
    {\pmb {\cal J}}^{\rm EH}_{s}(\boldsymbol{\rho})
    &=-\frac{ime\gamma_eE_0^2q_F^2}{8\hbar (2\pi)^3c}\int_{-\pi/2}^{\pi/2}d\theta_{\bf k}d{\theta_{\bf k'}}(\sin\theta_{\bf k'}\hat{\bf e}_\rho-\cos\theta_{\bf k'}\hat{\bf e}_\perp)\otimes\left[\hat{\bf e}_\rho(\cos\theta_{\bf k}+\cos\theta_{\bf k'})+\hat{\bf e}_\perp(\sin\theta_{\bf k}+\sin\theta_{\bf k'})\right]\nonumber\\
        &\times e^{iq_F\rho\cos\theta_{\bf k'}}e^{-iq_F\rho\cos\theta_{\bf k}}+{\rm H.c.}\nonumber\\
        &=-\frac{ime\gamma_eE_0^2q_F^2}{8\hbar (2\pi)^3c}(\hat{\bf e}_\rho\otimes\hat{\bf e}_\perp)\int_{-\pi/2}^{\pi/2}\sin^2\theta_{\bf k'}e^{iq_F\rho\cos\theta_{\bf k'}}d{\theta_{\bf k'}}\int_{-\pi/2}^{\pi/2}e^{-iq_F\rho\cos\theta_{\bf k}}d\theta_{\bf k}\nonumber\\
        &+\frac{ime\gamma_eE_0^2q_F^2}{8\hbar (2\pi)^3c}(\hat{\bf e}_\perp\otimes\hat{\bf e}_\rho)\left(\int_{-\pi/2}^{\pi/2}\cos^2\theta_{\bf k'}e^{iq_F\rho\cos\theta_{\bf k'}}d{\theta_{\bf k'}}\int_{-\pi/2}^{\pi/2}e^{-iq_F\rho\cos\theta_{\bf k}}d\theta_{\bf k}\right.\nonumber\\
        &+\left.\int_{-\pi/2}^{\pi/2}\cos\theta_{\bf k'}e^{iq_F\rho\cos\theta_{\bf k'}}d{\theta_{\bf k'}}\int_{-\pi/2}^{\pi/2}\cos\theta_{\bf k}e^{-iq_F\rho\cos\theta_{\bf k}}d\theta_{\bf k}\right)\nonumber\\
        &=\frac{me\gamma_eE_0^2q_F}{64\hbar\pi c\rho}\left((\hat{\bf e}_\perp\otimes\hat{\bf e}_\rho)+(\hat{\bf e}_\rho\otimes\hat{\bf e}_\perp)\right)\left(J_{0}(q_F\rho)H_{1}(q_F\rho)-J_{1}(q_F\rho)H_{0}(q_F\rho)\right),
\end{align}
where $J_{n}(x)$ and $H_{n}(x)$ are, respectively, the $n$-order Bessel function of the first kind and Struve function.

\section{Effect of chirality on transverse spin pumping}

\label{appendix_chirality}

As addressed in the main text, the unidirectionality of the Fourier components in the electromagnetic field is responsible for the asymmetrical distribution of the transverse spin current on the two sides of the magnetic wire. 
Here we expose more details by addressing its dependence on the ellipticity $\zeta^2$ of the ferromagnetic resonance, noting the field is unidirectional only when $\zeta^2=1$.
With the contour integral, the transverse spin current at the two sides of the magnetic wire reads
\begin{align}
    {\pmb{\cal J}}^{x}_y\left(x>\frac{w}{2}\right)&=\frac{e\mu_0^2\gamma_e}{32\pi^2 A}\sum_{{\bf q}}\sum_{\xi=\pm}\frac{q^2_y}{q^2_\xi}\hat{\bf y}\otimes\hat{\bf x}[f(\varepsilon_q)-f(\varepsilon_q-\xi\hbar\omega)](1-e^{-\abs{q_\xi-q_x}d})^2\nonumber\\
    &\times\frac{(\sin (q_\xi-q_x)w/2)^2}{(q_\xi-q_x)^3}(1+\xi\zeta^2{\rm sgn}(q_\xi-q_x))^2M_z^2{\rm sgn}(q_\xi-q_x)+{\rm H.c.},
    \label{Hall spin current+}\\
     {\pmb{\cal J}}^{x}_y\left(x<-\frac{w}{2}\right)&=\frac{e\mu_0^2\gamma_e}{32\pi^2 A}\sum_{{\bf q}}\sum_{\xi=\pm}\frac{q^2_y}{q^2_\xi}\hat{\bf y}\otimes\hat{\bf x}[f(\varepsilon_q)-f(\varepsilon_q-\xi\hbar\omega)](1-e^{-\abs{-q_\xi-q_x}d})^2\nonumber\\
    &\times\frac{(\sin (q_\xi+q_x)w/2)^2}{(-q_\xi-q_x)^3}(1+\xi\zeta^2{\rm sgn}(-q_\xi-q_x))^2M_z^2{\rm sgn}(-q_\xi-q_x)+{\rm H.c.}.
    \label{Hall spin current-}
\end{align}
\begin{itemize}
    \item With $\zeta^2=0$ by setting  $d\ll w$,  the unidirectionality of the electromagnetic field vanishes,  resulting in equal magnitudes of the transverse spin current on both sides:  ${\pmb{\cal J}}^x_y(x>w/2)|_{\zeta^2=0}={\pmb{\cal J}}^x_y(x<-w/2)|_{\zeta^2=0}$.
    \item With $\zeta^2=1$ by setting  $d=w$, the Fourier component of the field propagates undirectionally along the negative $\hat{\bf x}\mbox{-}$direction, breaking the reciprocity of the $``\xi=+"$ photon emission or $``\xi=-"$ photon absorption processes in the momentum space.  Particularly, the Hall spin current at the right-hand side of the magnet arises solely from the $\xi=``+"$ photon emission process, contributing to a transverse spin current [Eq.~\eqref{Hall spin current+}]
\begin{align} 
{\pmb{\cal J}}^{x}_y\left(x>\frac{w}{2}\right)=\frac{e\mu_0^2\gamma_e}{4\pi^2 A}\sum_{q_x<-q_\omega}\sum_{q_y}\frac{q^2_y}{q^2_+}\hat{\bf y}\otimes\hat{\bf x}[f(\varepsilon_q)-f(\varepsilon_q-\hbar\omega)](1-e^{-\abs{q_+-q_x}d})^2\frac{(\sin (q_+-q_x)w/2)^2}{(q_+-q_x)^3}M_z^2.
\nonumber
\end{align}
While at the left-hand side of the magnet, both the photon emission and absorption $\xi=\pm$  processes contribute to the Hall spin current according to Eq.~\eqref{Hall spin current-}.
\item In the intermediate regime with  $0<\zeta^2<1$, the electromagnetic field propagates bidirectionally. In this regime, both photon emission $``+"$ and absorption $``-"$  processes are allowed in the wave vector space. By varying $\zeta^2$, the relative scattering amplitudes of these processes can be controlled by the geometry of the magnetic wire, thereby tuning the relative magnitudes of the Hall spin currents generated in the left and right pumping regions. 
\end{itemize}

\end{widetext}

\end{appendix}

\end{document}